# The entropy functional, the information path functional's essentials and their connections to Kolmogorov's entropy, complexity and physics

Vladimir S. Lerner

13603 Marina Pointe Drive, Suite C-608, Marina Del Rey, CA 90292, USA, vslerner@yahoo.com

**Abstract**

The paper introduces the recent results related to an entropy functional on trajectories of a controlled diffusion process, and the information path functional (IPF), analyzing their connections to the Kolmogorov's entropy, complexity and the Lyapunov's characteristics. Considering the IPF's essentials and specifics, the paper studies the singularities of the IPF extremal equations and the created invariant relations, which both are useful for the solution of important mathematical and applied problems.

**Introduction**

The entropy functional, defined on a Markov diffusion process, plays an important roll in theory of information and statistical physics [1-3], informational macrodynamics and control systems [4].

However in the known references, we did not find the mathematical results related to the entropy functional's connections with the Kolmogorov's entropy, complexity, and the Lyapunov's characteristics [5]. We analyze these connections through the information path functional (IPF) of controllable diffusion process, considering the IPF's essentials and specifics, including the singularities of the IPF extremal equations and the created invariant relations. The paper is a part of the information path functional approach [6-8] in solving the problems of information evolutionary dynamics.

Searching a law, which governs *complex dynamics of interacting proce*ss, leads us to the *process' statistical dynamics* first, and then to finding a *variation principle* that, according to R. Feynman *(The character of physical law)*, might describe regularities of such dynamics, possibly at a macroscopic level. Sec.1 introduces the entropy functional expressed via an additive functional of controllable diffusion process and the parameters of corresponding stochastic equation, applied to the IPF variation problem. *Sec.*2. presents an *essence* of the IPF approach, considering both the problem statement and its formalization for a *class* of random systems, modeled by the solutions of controlled Ito's stochastic differential equations; defines a probabilistic measure of the functional's *distance* between a current *process' trajectory* and some *given process*; applies the *entropy functional*, defined on these solutions-trajectories, and presents the *dynamic approximation* of both this probabilistic measure and the conditional entropy functional by the corresponding *path functionals*. We *illustrate a specific* of the formulated variation problem's solution using both the Kolmogorov (K) eqs. for the functional of a Markov process and the Jacobi-Hamilton (J-H) eq. for the dynamic approximation of this functional as the IPF. Because the fulfillment of both J-H and K equations in the same field's region of a space is possible at some "punched" *discretely* selected points (DP), the extremal trajectory, solving the *variation* problem (VP) is divided on the extremal's *segments* at each DP.

Between the extremal segments (at DP) exists a "*window*", where the random information affects the dynamic process on the extremals, creating it's piece-wise *dependency* upon the observed data.

The solved VP allows the finding of both a *class* of the *dynamic (macro) models* (as the equation of the IPF extremals) for the considered *class* to the random systems (at the system's microlevel), and the *optimal control functions* for this *class of the dynamic models* (by solving the *optimal control's synthesis problem*). The synthesized optimal controls start at the beginning of each segment, act along the segment, and connect the segments in the macrodynamic optimal process, while the discrete interval of the applied control is associated with the segment's length between the DP. These specifics allow *providing the identification* of the model's dynamic operator at each DP for each extremal segment in *real time under* the optimal control action and during the object's current motion. Because the proofs of formulated theorems have published, we *illustrate here the theorems results.*

Sec.3 applies the obtained results for a joint solution of the optimal control and the identification problems, providing also the basic theorems for the creation of cooperative information dynamics.

In *sec*.4 we consider the IPF model's family of the interacting trajectories forming the complex system's *state consolidation and aggregation* in a *cooperative hierarchical information network* (IN). The IN's structure is based on the identified model's invariants, following from VP. The IN's formation can proceed concurrently during the system's optimal motion, combined with optimal control and the operator's identification.

Sec.5 studies the IPF macromodel's singular points and the singular trajectories, and the invariants following from their connections. The singularities arise at the DP-windows with shortening the initial model's dimension and the potential chaotic dynamics' bifurcations.

Sec.6 analyzes the solutions of a natural border problem for the IPF under the control actions. It is shown that both the model extremals and the model's singular trajectories belong to these solutions, if the segment's controls are bound by the found relations. We also establish the invariant conditions, as the model's field's functions, being the analogies of the information *conservations laws.*

Finally, in sec.7 we study the connections between the entropy's (information) path functional and the Kolmogorov's entropy of a dynamic system, between the Kolmogorov's and the macrodynamic complexities, and the relations to physics.

**1. The entropy functional**

Let have the $n$-dimensional controlled stochastic Ito differential equation [9]:

$$d\,\tilde{x}_t = a(t,\tilde{x}_t,u_t)dt + \sigma(t,\tilde{x}_t)d\,\xi_t,\ \tilde{x}_s = \eta,\ t \in [s,T] = \Delta,\ s \in [0,T] \subset R_+^1 \qquad (1.1)$$

with the standard limitations [9,10] on the functions of shift $a(t,\tilde{x}_t,u_t)$, diffusion $\sigma(t,\tilde{x}_t)$, and Wiener process $\xi_t = \xi(t,\omega)$, which are defined on a probability space of the elementary random events $\omega \in \Omega$ with the variables located in $R^n$; $\tilde{x}_t = \tilde{x}(t)$ is a diffusion process with transition probabilities $P(s,\tilde{x},t,B)$, and $\Psi(s,t)$ is a $\sigma$-algebra created by the events $\{\tilde{x}(\tau) \in B\}$, $s \le \tau \le t$; $P_{s,x} = P_{s,x}(A)$ are the corresponding conditional probability distributions on an extended $\Psi(s,\infty)$.

Let's consider the transformation of an initial process $\tilde{x}_t$, with transition probabilities $P(s,\tilde{x},t,B)$, to some diffusion process $\varsigma_t$, with transition probabilities

$$\tilde{P}(s,\varsigma_t,t,B) = \int_{\tilde{x}(t) \in B} \exp\{-\varphi_s^t(\omega)\} P_{s,x}(d\omega), \qquad (1.2)$$

where $\varphi_s^t = \varphi_s^t(\omega)$ is an additive functional of process $\tilde{x}_t = \tilde{x}(t)$ [11,12], measured regarding $\Psi(s,t)$ at any $s \le \tau \le t$ with probability 1, and $\varphi_s^t = \varphi_s^\tau + \varphi_\tau^t$.

Then at this transformation, the transitional probability functions $\tilde{P}(s,\varsigma_t,t,B)$ (1.2) determine the corresponding extensive distributions $\tilde{P}_{s,x} = \tilde{P}_{s,x}(A)$ on $\Psi(s,\infty)$ with the density measure

$$p(\omega) = \frac{\tilde{P}_{s,x}}{P_{s,x}} = \exp\{-\varphi_s^t(\omega)\}. \tag{1.3}$$

Using the definition of a conditional entropy [1] of process $\tilde{x}_t$ regarding process $\varsigma_t$:

$$S(\tilde{x}_t / \varsigma_t) = E_{s,x}\{-\ln[p(\omega)]\}, \tag{1.4}$$

where $E_{s,x}$ is a conditional mathematical expectation, we get

$$S(\tilde{x}_t / \varsigma_t) = E_{s,x}\{\varphi_s^t(\omega)\}. \tag{1.5}$$

Let the transformed process be $\varsigma_t = \int_s^t \sigma(v,\varsigma_v)d\varsigma_v$ having the same diffusion matrix as the initial process, but the zero drift. Then the above additive functional at its fixed upper limit $T$ acquires the form [11,12]:

$$\varphi_s^T = 1/2\int_s^T a^u(t,\tilde{x}_t)^T (2b(t,\tilde{x}_t))^{-1} a^u(t,\tilde{x}_t)dt + \int_s^T (\sigma(t,\tilde{x}_t)^{-1} a^u(t,\tilde{x}_t)d\xi(t),\ 2b(t,\tilde{x}) = \sigma(t,\tilde{x})\sigma^T(t,\tilde{x}) > 0, \tag{1.6}$$

where $E_{s,x}\{\int_s^T (\sigma(t,\tilde{x}_t)^{-1} a^u(t,\tilde{x}_t)d\xi(t)\} = 0$. \hfill (1.6a)

Finally we get the information entropy functional expressed via parameters of the initial controllable stochastic equation (1.1):

$$S(\tilde{x}_t / \varsigma_t) = 1/2 E_{s,x}\{\int_s^T a^u(t,\tilde{x}_t)^T (2b(t,\tilde{x}_t))^{-1} a^u(t,\tilde{x}_t)dt\}, \tag{1.7}$$

where for the variation problem in [4,6] relation $E_{s,x}[a^u(t,\tilde{x}_t)^T (2b(t,\tilde{x}_t))^{-1} a^u(t,\tilde{x}_t)] = E_{s,x}[\hat{L}(t,\tilde{x}_t)] = L$ plays a role of a Lagrangian $L = E_{s,x}[\hat{L}]$.

For a positive quadratic form in (1.7), the above information entropy is a positive.

*Example.* Let's have a single dimensional eq. (1.1) with the shift function $a^u = u(t)\tilde{x}(t)$ at the given control function $u_t = u(t)$, and the diffusion $\sigma = \sigma(t)$. Then the entropy functional has a view

$$S(\tilde{x}_t / \varsigma_t) = 1/2\int_s^T E_{s,x}[u^2(t)\tilde{x}^2(t)\sigma^{-2}(t)]dt, \tag{1.8a}$$

from which at the nonrandom $u(t), \sigma(t)$ we get

$$S(\tilde{x}_t / \varsigma_t) = 1/2\int_s^T [u^2(t)\sigma^{-2}(t)E_{s,x}[\tilde{x}^2(t)]]dt = 1/2\int_s^T u_t^2 \dot{r}_t^{-1} r_s dt, \tag{1.8b}$$

where for the diffusion process holds true: $2b(t) = \sigma(t)^2 = dr/dt = \dot{r}_t, E_{s,x}[\tilde{x}^2(t)] = r_s$, and the functional (1.8b) is expressed via the process' covariation functions $r_s$, $r_t$ and known $u_t$.

This allows us to *identify* the entropy functional on an observed controlled process $\tilde{x}_t = \tilde{x}(t)$ by measuring the above covariation (correlation) functions. The $n$-dimensional form of functional (1.8b) follows directly from using the related $n$-dimensional covariations and the control.

## 2. An essence of the information path functional approach

*The initial problem.* Let's have two random processes, one of them $\tilde{x}_t = \tilde{x}_t(u)$ is a controlled process, being a solution of eqs (1.1), another one $\tilde{x}_t^1$ is given as a programmed process, expressing a task for the controlled process (particularly, conveying a performance criterion); these could also be any two random processes.

*The problem formalization.* The control task can be formalize considering these process's $\delta-closeness$ by a probability measure $P[\rho_\Delta(\tilde{x}_t, \tilde{x}_t^1) < \delta]$, where $\rho_\Delta(\tilde{x}_t, \tilde{x}_t^1)$ is a metric distance in a Banach space, and *requiring*

$$SupP[\rho_\Delta(\tilde{x}_t, \tilde{x}_t^1) < \delta] \to SupP[\rho_\Delta(\tilde{x}_t^*, O) < \delta] \qquad (2.1)$$

or the closeness of the difference $\tilde{x}_t(u) - \tilde{x}_t^1 = \tilde{x}_t^*(u)$ to null-vector O. This problem we solve via the approximation of the random difference $\tilde{x}_t^*$ by a dynamic process $x_t$, which we call a *macroprocess*, defined in the space $KC^1$ (of the piece–wise-differentiable, continuous non random functions), while considering $\tilde{x}_t, \tilde{x}_t^1, \tilde{x}_t^*$ as the corresponding *microlevel* processes. Such a bi-level's micro- and macro- description we apply for a complex system with the above random processes at microlevel and dynamic disturbed processes at macrolevel with a disturbance process $\zeta_t$ for $x_t$. The process' description we concretize by modeling $\tilde{x}_t^*(u)$ by the solutions of a controlled stochastic differential equation Ito (analogous to (1.1)):

$$d\tilde{x}_t^* = a^u(t, \tilde{x}_t^*, u_t)dt + \sigma(t, \tilde{x}_t^*)d\xi_t, \tilde{x}_{t=s}^* = \tilde{x}_s^*, \qquad (2.2)$$

whose function of shift $a^u = a^u(t, \bullet, \bullet)$ is given, and the diffusion component of the solution, which models the disturbance $\zeta_t$ for $x_t$, has the same function of diffusion $\sigma = \sigma(t, \bullet, \bullet)$ as one in (1.1):

$$\zeta_t = \int_s^t \sigma(t, ., \upsilon)d\zeta_\upsilon \text{ at } E[\zeta_t] = O. \qquad (2.2a)$$

Control $u_t$ (in both (1.1) and (2.2)) is formed as a function of time and dynamic variables ($x_t$), defined by a feedback equation:

$$u_t \stackrel{def}{=} u(t, x_t), \qquad (2.2b)$$

where at a fixed $x \in R^n$, $u(\bullet, x)$ is a piece-wise continuous function of $t \in \Delta$, and at fixed $t \in \Delta$, $u(t, \bullet)$ is a continuous differentiable function, having the limited second derivatives by $x \in R^n$:

$$\forall x \in R^n, u(\bullet, x) \in KC(\Delta, U), u(\bullet, x) \in C^1(\Delta^o, U)KC, \forall t \in \Delta, u(t, \bullet) \in C^1(R^n, U), \qquad (2.2c)$$

being a piece-wise continuous function on $KC$, $C^1$ accordingly:

$$u_t \in KC(\Delta, U), u_+ \stackrel{def}{=} \lim_{t \to \tau_k + o} u(t, x_{\tau_k}), u_- \stackrel{def}{=} \lim_{t \to \tau_k - o} u(t, x_{\tau_k}), \Delta^o = \Delta \setminus \{\tau_k\}_{k=1}^m, k = 0, ..., m,$$

$$\tau_k \in \Delta, \tau_o = 0, \tau_m = T. \qquad (2.2d)$$

(We assume that the processes $\tilde{x}_t, \tilde{x}_t^1$ can also be modeled by the solutions of corresponding Ito stochastic equations, with the details in [13]). It was shown [14] that a Markov diffusion process is a convenient mathematical model for the representation of a wide class of many dimensional random processes.

Such processes and their nonstationary models in the class of stochastic differential equations and widely used in Statistical Physics, Irreversible Thermodynamics, Theory Information, and in controllable random processes theory [15,16, 2, 1,4, others].

The connection between the probabilities of above processes is expressed in the following form:

$$P[\rho_\Delta(\tilde{x}_t^*, O) < \delta] \geq P[\rho_\Delta(\tilde{x}_t^*, \zeta_t) < \delta]P[\rho_\Delta(\zeta_t, O) < \delta] \geq P[\rho_\Delta(\tilde{x}_t^*, x_t) < \delta]P[\rho_\Delta(\zeta_t, x_t) < \delta]P[\rho_\Delta(\zeta_t, O) < \delta] \qquad (2.3)$$

at $P[\rho_\Delta(\tilde{x}_t^*, \zeta_t) < \delta] \geq P[\rho_\Delta(\tilde{x}_t^*, x_t) < \delta] P[\rho_\Delta(\zeta_t, x_t) < \delta]$. (2.3a)

The proof, following from a triangle inequality and Markovian properties, is given in [13].
From that, the initial problem (2.1) can be reduced to the following requirements

$\underset{\tilde{x}_t^*}{Sup}\, P[\rho_\Delta(\tilde{x}_t^*, x_t) < \delta]$ (2.4a); $\underset{\tilde{x}_t^*}{Sup}\, P[\rho_\Delta(\tilde{x}_t^*, \zeta_t,) < \delta]$ (2.4b); $\underset{x_t}{Sup}\, P[\rho_\Delta(\zeta_t, x_t) < \delta]$ (2.4c); at $Sup\, P[\bullet] \leq 1$

where the last two conditions can be joint in the form

$\underset{\tilde{x}_t^*}{Sup}\, P[\rho_\Delta(\tilde{x}_t^*, \zeta_t,) < \delta] \Rightarrow \underset{x_t}{Sup}\, P[\rho_\Delta(\zeta_t, x_t) < \delta]$ . (2.4d)

Relation (2.4a) represents a probability condition of the identification $\tilde{x}_t^*$ via $x_t$, while the pair in (2.4d) minimizes the deviation $\tilde{x}_t^*$ from $\zeta_t$ through a minimal deviation of $x_t$ (as a macromodel of $\tilde{x}_t^*$) from $\zeta_t$ and connects them. We assume here that process $\zeta_t$ also models an irremovable disturbance for $x_t$. This is why for $x_t$, to be a dynamic analog of $\tilde{x}_t^*$, we also require the fulfillment of (2.4c) and the connection of both probabilistic closeness's by a mutual ability to approximate $\zeta_t$ (considered as a standard process by (2.4d)). According to this condition, the *control*, moving the difference $\tilde{x}_t^*$ close to $\zeta_t$, also approximates $x_t$ with accuracy of $\zeta_t$ and, therefore, leads to the approximation of $\tilde{x}_t^*$ by $x_t$, which *redefines the initial control problem*.

For the evaluation of above probabilities' conditions we use the Freidlin-Wentzel results [17] applying them to the considered control system in the form

$$\lim_{\varepsilon \downarrow 0} \sup \varepsilon \log P[\rho_\Delta(\tilde{x}_t^*, x_t) < \delta] \leq - \underset{x_t}{Inf}\, S_1(x_t),$$ (2.5)

where $S_1(x_t) = 1/2 \int_s^T (\dot{x}_t - a^u(x_t))^T (2b_t)^{-1} (\dot{x}_t - a^u(x_t)) dt$ (2.5a)

is a *path functional* along trajectory $x_t$, which approximates the difference $\tilde{x}_t - \tilde{x}_t^1 = \tilde{x}_t^*$ with a maximal probability measure (2.4a); and

$$\lim_{\varepsilon \downarrow 0} Sup\, \varepsilon \log P[\rho_\Delta(\zeta_t, x_t) < \delta] \leq - \underset{x_t}{Inf}\, S_2(x_t),$$ (2.5b)

where $S_2(x_t) = 1/2 \int_s^T \dot{x}_t^T (2b_t)^{-1} \dot{x}_t dt$ (2.5c)

is *a path functional* which evaluates the deviation of the trajectory $x_t$ from $\zeta_t$. We get also

$$\lim_{\varepsilon \downarrow 0} Sup\, \varepsilon \ln P[\rho_\Delta(\tilde{x}_t^*, \zeta_t) < \delta] \leq - \underset{\tilde{x}_t^*}{Inf}\, S_3(\tilde{x}_t^*),$$ (2.6)

where the probability's logarithm of the transformation $\tilde{x}_t^*$ to $\zeta_t$ can be written through the Radom-Nikodim density measures [12,1] $\dfrac{d\mu^\zeta}{d\mu^{\tilde{x}_t^*}}$ of the above functions on a set $B_\delta$:

$\ln P[\rho_\Delta(\tilde{x}_t^*, \zeta_t) < \delta] = \int_{B_\delta} \ln \dfrac{d\mu^\zeta}{d\mu^{\tilde{x}_t^*}}(\tilde{x}_t^*) P(d\tilde{x}_t^*) = E[\ln \dfrac{d\mu^\zeta}{d\mu^{\tilde{x}_t^*}}]$, $B_\delta = \{\rho_\Delta(\tilde{x}_t^*, \zeta_t) < \delta\}$ (2.7)

and $E[\ln \dfrac{d\mu^\zeta}{d\mu^{\tilde{x}_t^*}}] = E[-\ln \dfrac{d\mu^{\tilde{x}_t^*}}{d\mu^\zeta}] = -S(x_t^*/\zeta_t)$ , (2.7a)

defines a *conditional entropy* $S(x_t^*/\zeta_t)$ of processes $\tilde{x}_t^*$ regarding $\zeta_t$ (which we connect below to functional $S_3(\tilde{x}_t^*)$). For a Markov diffusion process, the density measure is expressed through an additive functional $\varphi_s^T$ of considered diffusion processes [11, 12], sec.1:

$$\frac{d\mu^\zeta}{d\mu^{\tilde{x}_t}} = \exp(-\varphi_s^T) = \exp[-(\tilde{S}_3 + \int_s^T ((\sigma(t,\tilde{x}_t^*))^{-1} a^u(t,\tilde{x}_t^*) \, d\zeta_t)], \; \tilde{S}_3 = \tilde{S}_3(\tilde{x}_t^*), \tag{2.8a}$$

$$\tilde{S}_3 = 1/2 \int_s^T a^u(t,\tilde{x}_t^*)^T (2b(t,\tilde{x}_t^*))^{-1} a^u(t,\tilde{x}_t^*) dt, \; 2b = \sigma\sigma^T; \tag{2.8b}$$

and the entropy (2.7a) is defined via the additive functional (2.8b) in the form [1]:

$$S(x_t^*/\zeta_t) = E[\tilde{S}_3], \tag{2.9}$$

at $E[\int_s^T ((\sigma(t,\tilde{x}_t^*))^{-1} a^u(t,\tilde{x}_t^*) d\zeta_t] = 0$. (2.9a)

Thus, relation (2.7) is defined by the conditional entropy of processes $\tilde{x}_t^*$ regarding $\zeta_t$:

$$Sup \ln P[\rho_\Delta(\tilde{x}_t^*, \zeta_t) < \delta] = Sup - S(\tilde{x}_t^*/\zeta_t) = Inf S(\tilde{x}_t^*/\zeta_t) = Inf\{E[\tilde{S}_3]\}, \tag{2.10}$$

and according to (2.6), the lower entropy level is limited by:

$$Inf S(\tilde{x}_t^*/\zeta_t) \leq -Inf S_3(\tilde{x}_t^*), \; S_3(\tilde{x}_t^*) = 1/2\int_s^T a^u(t,\tilde{x}_t^*) 2b(t,\tilde{x}_t^*)^{-1} a^u(t,\tilde{x}_t^*) dt. \tag{2.10a}$$

Finally we come to the variation conditions

$$\underset{x_t}{Inf} \, S_1(x_t), \tag{2.11a}$$

$$\underset{x_t^*}{Inf} \, S(x_t^*/\zeta_t) \Rightarrow \underset{x_t}{Inf} \, S_2(x_t) \tag{2.11b}$$

whose fulfillments solve jointly the above problems of *optimal control and the identification* and determines the *macroprocess as an extremal* of the variation problem.

Relation (2.11b) also connects the path functional approach to theory information and allows a dynamic approximation of the entropy functional of diffusion process by the *information* path functional (IPF).

The *specific* of *solution* of the above variation problem (VP) we illustrate using condition (2.11b) in the form: $\min S_2 = \min S_3 = \min E(\tilde{S}_3)$, (2.12)

$$S_2 = \int_s^T L(t,x,\dot{x}) dt, L(t,x,\dot{x}) = 1/2\dot{x}^T (2b)^{-1} \dot{x}, \tag{2.13}$$

$$\tilde{S}_3 = \int_s^T L(t,x,a^u) dt, \; L(t,x,a^u) = 1/2(a^u)^T (2b)^{-1} a^u. \tag{2.13a}$$

The Jacobi-Hamilton (J-H) equation [18] for the extremals $x_t = x(t)$ of functional (1.13) is

$$-\frac{\partial S_2}{\partial t} = H, \; H = \dot{x}^T X - L(t,x,\dot{x}), \tag{2.13b}$$

where $X$ is a conjugate vector for $x$. Using the Kolmogorov (K) equation [5,12, 19, others] for the functional (2.13a) in a field for the Markov diffusion process (from (2.2)) we get

$$-\frac{\partial \tilde{S}_3}{\partial t} = a^u \frac{\partial \tilde{S}_3}{\partial x} + b \frac{\partial^2 \tilde{S}_3}{\partial x^2} + 1/2(a^u)^T (2b)^{-1} a^u. \tag{2.13c}$$

According to condition (2.12) we require

$$-\frac{\partial S_2}{\partial t} = -\frac{\partial \tilde{S}_3}{\partial t}, \frac{\partial S_2}{\partial x} = \frac{\partial \tilde{S}_3}{\partial x} = X, \qquad (2.14)$$

which leads to

$$-\frac{\partial \tilde{S}_3}{\partial t} = a^u X + b\frac{\partial X}{\partial x} + 1/2(a^u)^T (2b)^{-1} a^u. \qquad (2.14a)$$

The fulfillment of both J-H and K equations in the same field's region of a space is possible at some "punched" *discretely* selected points (DP) of the space

$$R^n : \Gamma_\varphi = \bigcup_{i=1}^{m} \tau_i, i = 1,...,m, \qquad (2.14b)$$

where the field of functional (2.13) can coincide with the field of entropy functional (2.13a), defined on the microlevel's diffusion process. Applying to (2.14a) the Hamiltonian in eq.(2.13b) we have

$$\frac{\partial(-\frac{\partial S_2}{\partial t})}{\partial X} = \frac{\partial H}{\partial X} = \dot{x}, \; \frac{\partial(-\frac{\partial S_2}{\partial t})}{\partial X} = \frac{\partial(-\frac{\partial \tilde{S}_3}{\partial t})}{\partial X} = a^u, \dot{x} = a^u. \qquad (2.15)$$

The equation for the conjugate vector we get using the Lagrange eq. for $L(t, x, \dot{x}) = 1/2\dot{x}^T (2b)^{-1} \dot{x}$:

$$X = \frac{\partial L}{\partial \dot{x}} = (2b)^{-1} \dot{x}. \qquad (2.16)$$

The substitution (2.16) to (2.13b) brings the Hamiltonian (2.13b) to the form

$$H = 1/2\dot{x}^T X = 1/2\dot{x}^T (2b)^{-1} x, \qquad (2.17)$$

and according to (2.14), the equalization of both relations (2.17) and (2.14a):

$$-\frac{\partial \tilde{S}_3}{\partial t} = a^u X + b\frac{\partial X}{\partial x} + 1/2(\dot{x})^T (2b)^{-1} \dot{x} = -\frac{\partial S_2}{\partial t} = 1/2(\dot{x})^T (2b)^{-1} \dot{x} \qquad (2.18)$$

determines the eqs. of a *constraint* imposed by the microolevel's stochastics (according to (2.13a)):

$$a^u(\tau)X(\tau) + b(\tau)\frac{\partial X}{\partial x}(\tau) = 0, \qquad (2.18a)$$

at which the J-H and K equations coincide at DP (2.14b). After substituting $a^u = 2bX$ (following from (2.15, 2.16)) at $b \neq 0$, the constraint acquires the forms

$$\frac{\partial X}{\partial x}(\tau) = -2X(\tau)X^T(\tau) \; (2.18b); \text{ or } a^u(\tau) = \sigma(\tau)\sigma(\tau)^T X(\tau). \qquad (2.18c)$$

From (2.14b), (2.18) follow that the constraint equations (2.18b,c), which establish a connection between the microlevel's diffusion and macrolevel's dynamics, can be relevant only at these discrete points (DP), while macroequation $\dot{x} = a^u$ acts along each extremals except the punched points. This constrain allocates a set of the discrete states $x_\tau = x(\tau)$ for which IPF coincides with the entropy functional.

The constrain corresponds to the operator's equation

$$\tilde{L}S[x_\tau] = 0, \tilde{L} = a^u \frac{\partial}{\partial x} + b\frac{\partial^2}{\partial x^2}, \text{ at } \Delta S(x_\tau) = S(x(\tau_k)) - S(x(\tau_{k+1})) = inv, \qquad (2.18d)$$

whose solutions allow classifying the punched points, considered to be the bordered points of a diffusion process $\lim_{t \to \tau} \tilde{x}(t) = x(\tau)$ [12]. A bordered point $x_\tau = x(\tau)$ is attracting the only if function

$$R(x) = \exp\{-\int_{x_o}^{x} a^u(y)b^{-1}(y)dy\}, \qquad (2.18e)$$

defining the general solutions of (2.18d), is integrable at a locality $x = x_\tau$, satisfying the condition

$$|\int_{x_o}^{x_\tau} R(x)dx| < \infty .$$  (2.18f)

Using (2.15,2.16) we may write (2.18e) in the form

$$R(x) = \exp(-2\int_{x_o}^{x} X(y)dy).$$  (2.18g)

A bordered point is repelled if the eq.(2.18d) does not have the limited solutions at this locality, means that the above eqs is not integrable. The eqs. (2.18e,f) are the necessary and sufficient conditions of an existence of the solutions (2.18a-c), which define a set of the states $x_\tau = x(\tau)$, where the macrodynamics arise from stochastics, and determine some boundary conditions, limiting the above set. The necessary condition for the punched points to be attractive: $b(y) \neq 0$, corresponds existence of a regular diffusion process [12], and $a^u(y) \neq 0$ determines a potential creation of the dynamics; at $b(y) = 0$ both the entropy functional and IPF are degenerated: $S_3 \to \infty, S_2 \to \infty$; at $a^u(y) = 0$ the process' dynamics are vanished. Therefore, the fulfillment of (2.18ef) guarantees that eq. (2.18d) is integrated, the punched points exist and are attractive, where the dynamics can start. This *brings a quantum character of generation for both the macrostates and the macrodynamic's information at the VP fulfillment.*

A total information, *originated by the macrodynamics*, is equal to:

$$S(\Gamma_\tau) = 1/2 \int_{\Gamma_\tau} (a^u(\tau_i))^T (2b(\tau_i))^{-1} a^u(\tau_i) d\tau_i, \Gamma_\tau = \bigcup_{i=1}^{m} \tau_i, i = 1,...,m ,$$  (2.19a)

where the operator shift and the diffusion matrix are limited by eqs. (1.18a,b), $\Gamma_\tau$ is an union of a total number of $\tau_i$ time's instants for *n*-dimensional model (1.1). Eq. (1.20) also satisfies a *stationary* condition at a $\tau_i$-locality.

The identified DP divide the macrotrajectory on a *sequence* of the extremal segments limited by the punched localities, where the model's *randomness and regularities are connected*, and therefore the model's identification is possible. At these points, the constraint (2.18b) is applicable for the identification in the form

$$E[2X(\tau)X^T(\tau) + \frac{\partial X}{\partial x}(\tau)] = 0.$$  (2.19)

Writing the equation of extremals $\dot{x} = a^u$ in a traditional form:
$$\dot{x} = Ax + u, u = Av, \dot{x} = A(x+v),$$  (2.20)
where $v$ is a control reduced to the stare vector $x$, we will identify matrix $A$ and find the control $v$ that solve the initial problem. Substituting

$$X = (2b)^{-1} A(x+v), X^T = (x+v)^T A^T [(2b)^{-1}]^T, \frac{\partial X}{\partial x} = (2b)^{-1} Ax$$  (2.21)

into (2.19) we get $(2b)^{-1} A = -2E\{(2b)^{-1} A(x+v)(x+u)^T A^T (2b)^{-1}\}$,  (2.22)
from which at a nonrandom $A$ and $E[b] = b$, we obtain the eqs for the identification of
$$A(\tau) = -b(\tau)r_v^{-1}(\tau), r_v = E[(x+v)(x+v)^T], b = 1/2\dot{r}, r = E[\tilde{x}\tilde{x}^T]$$  (2.23)
via the above correlation functions. The results, which formalize the *object's dynamic macromodel* and the synthesized *optimal controls*, follow from Theorems (T1, T2) below (proved in [6]).

*Theorem 1(T1)*. *The equations for both functional fields (defined by K and J-H) of the VP are satisfied jointly at a limited set $\Delta^o$ where the following equations for the macromodel and controls hold true:*

$\dot{x} = a^u$, $a^u = A(t,x)x + u$, $u = A(t,x)v$, $A(t,x) = A \in KC(\Delta, L(R^n)) \cap C^1(\Delta^o, L(R^n))$, $(t,x) \in (\Delta \times R^n)$, (2.24)

$$\Delta^o = \Delta \setminus \Gamma_\varphi, \Gamma_\varphi = \bigcup_{k=1}^{m} \tau_k, \, v_t \in KC(\Delta, V) \cap C^1(\Delta^o, L(R^n)), V \subset R^n, \quad (2.25)$$

*where $v = A^{-1}u$ is the control vector $u$, reduced to a state vector $x$, with rank[v]=rank[x]=n, A(t,x) is a nonsingular macromodel's matrix; $\Gamma_\varphi$ is a "punched" set of a discrete points (DP) $\tau \in \bigcup_{k=1}^{m} \tau_k$ in $\Delta$; $C^1$ and KC accordingly are the space of the continuous differentiable and the piece-wise differentiable on $\Delta$, n-dimensional vector-functions.* •

This means that the DP divide the macrotrajectory on a sequence of the extremal's segments, defined by the solutions of macromodel (2.24), while the controls (2.25) are applied at a beginning of each segments. These extremals provide a piece-wise approximation of the initial entropy functional with the aid of the controls.

*Theorem 2* (T2). The VP is solved under

(1)- the class of the *piece-wise controls* (2.25) being fixed at each segment;

(2)-the controls which are *switched* at the DP $\tau \in \bigcup_{k=1}^{m} \tau_k$, defined by the conditions of equalization of the *dynamic model relative phase speeds*:

$|dx_i / dt \, (\tau_k - o) \, x_i^{-1}(\tau_k)| = |dx_j / dt \, (\tau_k - o) \, x_j^{-1}(\tau_k)|, x_i(\tau_k) \neq 0, x_j(\tau_k) \neq 0, \, i,j = 1,...,n;$ (2.26)

(3)-the controls, which at moments (2.26) *change* the model's matrix from $A_- = A(\tau_{k-o})$ to its *renovated* form $A_+ = A(\tau_k)$ (at a subsequent extremal segment), while both matrices are identifiable by the following relations for the conditional covariance (correlation) functions:

$A_- = 1/2 \, \dot{r} r^{-1} = 1/2 \, r^{-1} \dot{r}_-, \dot{r}_- = \dot{r}(\tau_k - o), r = E_-(xx^T), E_- = E_{\tau_{k-o}}, \dot{r}(\tau_k) = 2b(\tau_k),$ (2.27)

$A_+ = \pm A_-(1 + \mu_v^1)^{-1} = \pm (1 + \mu_v^1)^{-1} A_-, \mu_v^1 \in R^1, \mu_v^1 \neq -1, or$ (2.27a)

$A_+ = \pm 1/2 \, r^{-1} \dot{r}_- (1 + \mu_v^2)^{-1} = \pm 1/2 \, \dot{r}_- r^{-1} (1 + \mu_v^2)^{-1}, \mu_v^2 \in R^1; \mu_v^2 \neq -1;$ (2.27b)

(4)- the *control function*

$v_- = \angle_v^1 x_-, v_- = v(\tau_k - o), x_- = x(\tau_k - o), \angle_v^1 = \mu_v^1 I \neq 0,$ (2.28)

which changes *matrix $A_-$ to $A_+$* (according to (2.27a)), and the *control function*:

$v_+ = \angle_v^2 x_+, \angle_v^2 = \mu_v^2 I \neq 0, x_+ = x(\tau_k),$ (2.28a)

which changes matrix $A_+$ to $A(\tau_k + o) = \pm A(\tau_k)(I + \angle_v^2)$ (according to (2.27b), *or the control function*:

$v_+ - v_- = \angle_v^1 (x_- + v_-), v_+ = v(\tau_k)$, which changes the above matrix to $A(\tau_k + o) = \pm A(\tau_k)(I + \angle_v^1)$,

where coefficients $\mu_v^2 = (0, -2)$ satisfy applying the feedback control, which fulfills $A_+ = -A_-$, and brings the control functions (2.28, 2.28a,b) to the forms

$v_- = -2x(\tau - o), v_+ = -2x(\tau),$ (2.29)

$\delta v = v_+ - v_- = -2x(\tau) - v_-, v_- = -2x(\tau - o), \delta v(o) = v_t^\delta, o = o(\tau_k) = (\tau_{k-o}, \tau_k).$ (2.30) •

*Comments.* The last equation determines the control jump (a "needle" control's $v_t^\delta$ action), which connects the subsequent extremal segments. Controls (2.29, 2.30), solving the VP, we call the *optimal* controls, which start at the beginning of each segment, act along the segment, and connect the segments in the macrodynamic optimal process. The needle $\delta$-control, acting between the moments $(\tau_{k-o}, \tau_k)$, also performs a decoupling (a "decorrelation") of the pair correlations at *these* moments. The reduced control presents a projection of control $u_t$ on each of the state macrocoordinates, which is consistent with the object's controllability and identifiability [20, other]. This control *specifies* the structure of the controllable drift-vector $a^u = A(x+v)$ and the model's dynamic operator, which is identifiable using the *identification equations* (2.27, 2.27a) for the correlations functions, or the equation identifying directly the operator:

$$A(\tau) = b(\tau)(2\int_{\tau-o}^{\tau} b(t)dt)^{-1} > 0 \tag{2.31}$$

by the dispersion matrix *b* from (2.2a, 2.8b).

The control provides also the fulfillment of equality (2.26), which identifies each following DP.

The reduced controls, built by the macrostates that are *memorized* at $(\tau_{k-o}, \tau_k)$, according to (2.19),(2.20), are an important part of the macrosystem's structure, providing a mechanism of a self-control synthesis. These controls are also applied for a direct programming and the process' prognosis.

Let's *illustrate* the theorem's results, considering alongside with model (2.20) the model of a closed system (with a negative feedback) in the form $\dot{x}_t(t) = -A^v(t)x(t)$, where matrix $A^v(t)$ is a subject of both its definition and identification. Using relations $X = (2b)^{-1}(-A^v x)$ and constraint (2.19), we get

$$A^v(\tau) = b(\tau)r_x^{-1}(\tau), r_x = E[xx^T]. \tag{2.32}$$

Both forms for $a^u(\tau) = A(\tau)(x(\tau) + v(\tau)) = -A^v x(\tau)$ (2.32a)

and the identification eqs (2.23), (2.32) for $A(\tau)$ and $A^v(\tau)$ coincide at $v(\tau) = -2x(\tau)$, while it is also fulfilled

$$r_x = r = E[\tilde{x}\tilde{x}^T], b = 1/2\dot{r}, \tag{2.33}$$

where $r = r(\tau)$ is a covariation matrix, determined at the $o(\varepsilon) \sim \delta\tau$-locality (connected the micro-and macrostates). Using the equivalent equations: $\dot{x} = 2bX, \dot{x} = -br^{-1}x$, we get expression for

$$X = -1/2hx, h = r^{-1}. \tag{2.33a}$$

and

$$X(\tau) = -1/2(\int_{\tau-0}^{\tau} \sigma\sigma^T dt)^{-1}x(\tau). \tag{2.33b}$$

A potential, corresponding to the conjugate vector, which satisfies eq.(2.16) at the DP, loses its *deterministic* dependency on the shift vector (2.2), becoming the function of *diffusion* and a state vector at the DP vicinity (2.33b). The gradient in (2.33b) depends only on the diffusion:

$$gradX(\tau) = \frac{\partial X(\tau)}{\partial x(\tau)} = -1/2(\int_{\tau-0}^{\tau} \sigma\sigma^T dt)^{-1} = -2X(\tau)X^T(\tau) \tag{2.33c}$$

and at the vicinity's border, where $\sigma\sigma^T \to 0$, it acquires a form of the $\delta$-function. Out of the DP, the gradient (2.33c) does not exist as well as the potential function in the form (2.33b). The kinetic form for the conjugate vector still satisfies (2.18), where the kinetic operator is determined by its macroscopic value in (2.18c). Thus, the equalities (2.18a,b,c), (2.26), and (2.33a,b,c) (following from (2.12)) define a set of states $X(\tau), x(\tau)$ on the extremal trajectory, which are used for an access to the random process, specifically by forming the control functions (2.29,2.30) (where the controls are a part of the shift vector in (2.2)) and the operator identification (2.31, 2.32). (For example, from (2.18c, 2.33) follows

$$x(\tau) = 2r(\tau)\dot{r}^{-1}(\tau)a^u(\tau).\tag{2.33d}$$

From other consideration, using (2.35),(2.18b), and (2.15,2.16), we get direct connection the shift vector and diffusion in the form $a^u(\tau)(a^u(\tau))^T = b(\tau)(\int_{\tau-0}^{\tau} 2b(t)dt)^{-1}b(\tau)^T$. (2.33e)

(The last relation coincides with (2.32, 2.32a) at $x(\tau)x(\tau)^T = r(\tau)$).

For $a^u(\tau) = -A(\tau)x(\tau) = b(\tau)r^{-1}(\tau)x(\tau)$, function $R(x) = \exp\{\int_{x_o}^{x} r^{-1}(y)y\,dy\} < \infty, y = x(\tau) \neq 0$ if $r(y) \neq 0$, which is satisfied for a regular diffusion process. Substituting in (2.33c),(2.19) relations $\int_{\tau-0}^{\tau} \sigma\sigma^T dt = r(\tau), b(\tau) = 1/2\dot{r}(\tau), A(\tau) = -b(\tau)r^{-1}(\tau)$, we get

$$A(\tau) = -2E[\dot{x}(\tau)X^T(\tau)], \text{where } E[\dot{x}_i(\tau)X_k(\tau)] = E[\dot{x}_i(\tau)X_i(\tau)] = 2E[H_i] = -2E[\frac{\partial S_i}{\partial t}],\tag{2.33f}$$

following from
$E[\dot{x}_i(\tau)X_k(\tau)] = E[\lambda_i(\tau)x_i(\tau)(-1/2)h_{ik}(\tau)x_k(\tau)] = -1/2\lambda_i(\tau), E[x_i(\tau)_{ik}x_k(\tau)h_{ik}(\tau)] = 1$,
$E[\dot{x}_i(\tau)X_i(\tau)] = E[\lambda_i(\tau)x_i(\tau)(-1/2)h_{ii}(\tau)x_i(\tau)] = -1/2\lambda_i(\tau), E[x_i(\tau)_{ii}x_i(\tau)h_{ii}(\tau)] = 1$.

We come to $\lambda_i(\tau) = 4E[\frac{\partial S_i}{\partial t}(\tau)], \frac{\partial S_i}{\partial t} = \frac{\partial \tilde{S}_3^i}{\partial t} = \frac{\partial S_2^i}{\partial t}$, (2.33g)

which establishes the eigenvalue's connection to the above local differential entropy, taken at $t = \tau$ along the IPF for each $i$-model's dimension ($i=1,…n$). The above math expectation brings an average differential entropy for each dimension. If each $i$-dimension contains $k$ extremal segments, then $\lambda_{ij}$ indicates the $i$-th eigenvalue of $j$-th segment and $\lambda_{ij}(\tau) = 4E_j[\frac{\partial S_{ij}}{\partial t}(\tau)]$ presents the math expectation for each $k$-th segment at its $t = \tau$. The differential entropy's sum for all k-segments:

$$\sum_{j=1}^{k} E_j[-\frac{\partial S_{ij}}{\partial t}(\tau)] = \sum_{j=1}^{k} \lambda_{ij}(\tau) = TrA(\tau) \text{ equals to } E[\frac{\partial \tilde{S}_3}{\partial t}(\tau)].$$

Applying the optimal controls (2.29,2.30) to the invariant relation (2.18d)(right) brings the fllowing invariants:

$$\lambda_i(\tau_k)\Delta t_k = inv, \lambda_i(\tau_k)\tau_k = inv; \lambda_i(\tau_{k+1})\tau_k = inv,\tag{2.34}$$

where $\Delta t_k = \tau_{k+1} - \tau_k$ is a time interval between DP: $\tau = (\tau_1, \tau_2, ..., \tau_{k-1}, \tau_k,...), \lambda_i(\tau_k)$ is the eigenvalue taken at the moment $\tau_k$. Because the constraint (2.19) acts at each DP moment $\tau = (\tau_1, \tau_2, ..., \tau_{k-1}, \tau_k,...)$ for a sequence of the extremal segments, the controls at the nearest moments: $v(\tau_k) = -2x(\tau_k), v(\tau_{k-1}) = -2x(\tau_{k-1})$, where $\tau_k = \tau_{k-1} + \Delta t_k$. And after applying the last of the control to eq. (2.23) and substituting its solution at $\Delta t_k$ to (2.32a), we get the connection of the above matrices at any $\Delta t_{k-1}, \Delta t_k$:

$$A^v(\Delta t_k) = A(\tau_{k-1})\exp[(A(\tau_{k-1})\delta\tau_k]\{2-\exp[(A(\tau_{k-1})\Delta t_k]\}^{-1},\tag{2.35}$$

where $A(\tau_{k-1})$, being identified at the moment $\tau_{k-1}$, also determines $\Delta t_k$ from (2.34, 2.31).

The identification at the moment $\tau_k$ brings the new or renovated $A(\tau_k), A^v(\tau_k)$, and so on. In the procedure of the matrix identification [19], interval $\Delta t_k$ is used for the matrix's computation from the data had obtained at each DP $\tau_{k-1}$. The above discrete control, applied at beginning of $\Delta t_k$ proceeds during this interval, while at a moment $\delta\tau_k$ between the segments the needle control $\delta v$ (2.30) is applied, which connects the extremal's segments. From the variation eqs. (2.18), (2.17) follow that on the extremals holds true the condition $\min E[-\frac{\partial \tilde{S}_3}{\partial t}(\tau)] = \min H(\tau)$.

Applying the last one in the form $\min E[H(t)]$ and substituting in it (2.20, 2.21, 2.23, 233e) at the above optimal control, we come to the condition

$$\min E[H(\tau)] = \max E[\frac{\partial \tilde{S}_3}{\partial t}] = \min Tr[A(\tau-o)], Sign A(\tau-o) = Sign A(\tau+o), \qquad (2.36)$$

which connects the the identified matrix's elements to the initial variation conditions (2.12) and the above eqs.

## 3. The IPF results for a joint solution of the optimal control, identification, and the consolidation's problems

Solution of this *problems* we consider for a system, observed *discretely* at the moments of applying the optimal control: $\tau \in \{\tau_k\}$, $k = 1,...,m$, and transformed by this control to the terminal state $x_T = 0$.

Let us apply a transformation $G$ to the model (2.24), transforming it to a diagonal form:

$$dz/dt = \bar{A}(z+v), \bar{A} = G^{-1}AG, G = (G_{ij}) \in L(R^n), \det G \neq 0, \forall t \in \Delta^\circ, z = Gx, \bar{v} = Gv, \qquad (3.1)$$

$$x_T = (o_{ij})_{i,j=1}^n = O \Leftrightarrow z_T = (o_{ij})_{i,j=1}^n = O, \bar{v} = -2z(\tau), \bar{A}^v = \bar{A}(I + (\frac{\bar{v}_j(\tau,\cdot)}{z_i(t,\cdot)}\delta_{ij})_{i,j=1}^n = (\lambda_i(t))_{i=1}^n, \qquad (3.2)$$

where the piece-wise matrices $A, \bar{A}$ are fixed within the intervals of the control discretization $t_k, k = 1,...,m-1$, and are identifiable at each of these intervals, while the matrices eigenvalues (3.2) are connected according to relations (2.26); $I$ is identity matrix.

*Theorem* 3.1 (T3.1).

Transferring the system (3.1) to an origin of its coordinate system by the optimal controls, applied at the time intervals $t_k, k = 1,...,m$, requires the existence of a minimum of *two* matrix's $\bar{A}^v = (\lambda_i^k)_{i=1}^n$ eigenvalues, which at each of these moments satisfy the condition of connecting these intervals in the forms

$$|\lambda_i^k| = |\lambda_j^k|, i, j = 1,...,n, k = 1,...,m-1 \qquad (3.3)$$

with the *number of the control discrete intervals equal to n*.

*Proof.* By applying (2.26) to (3.1), using the matrix function (2.35) under the control $\bar{v} = -2z(\tau_{k-1})$, we come to the recurrent relations connecting the nearest $\lambda_i^k, \lambda_i^{k-1}$:

$$\lambda_i^k = -\lambda_i^{k-1}\exp(\lambda_i^{k-1}t_k)(2-\exp(\lambda_i^{k-1}t_k))^{-1}. \qquad (3.4)$$

Then solutions of (3.2) acquire the form

$$z_i^k(t_k) = (2-\exp(\lambda_i^{k-1}t_k)z_i^{k-1}(t_{k-1}). \qquad (3.5)$$

By writing the solution on the last control's discrete interval $t_m = T$:

$$z_i(T,\cdot) = (2-\exp(\lambda_i^{m-1}T)z_i(t_{m-1}) = 0, z_i(t_{m-1}) \neq 0, i == 1,...,n, \qquad (3.6)$$

we get the relation, defining $T$ through a preceding eigenvalue, which satisfies to all previous equalizations:

$$T = t_{m-1} + \ln 2/|\lambda_i^{m-1}|, \lambda_1^{m-1} > 0, \lambda_1^{m-1} = \lambda_2^{m-1}....= \lambda_n^{m-1} > 0. \qquad (3.7)$$

The positivity of the above eigenvalues can be reached at applying the needle controls in addition to the above step-wise controls. If these controls are not added, more general conditions below are used.
The equalizations of the eigenvalues at other discrete intervals, leads to the chain of the equalities for $n \geq m$:

$$|\lambda_1^{m-1}|=|\lambda_2^{m-1}|=....=|\lambda_n^{m-1}| \quad (3.8)$$

$$|\lambda_1^{m-2}|=|\lambda_2^{m-2}|=.....=|\lambda_{n-1}^{m-2}|,...,$$

$$|\lambda_1^{m-i-1}|=|\lambda_2^{m-i-1}|=....=|\lambda_{n-i}^{m-i-1}|,...,$$

$$|\lambda_1^1|=|\lambda_2^1|=.....=|\lambda_{n-m+2}^1|, \quad (3.8a)$$

and for $m \geq n$ leads to the following chain of the equalities:

$$|\lambda_1^{m-1}|=|\lambda_2^{m-1}|....=|\lambda_n^{m-1}|, (3.9)$$

$$|\lambda_1^{m-2}|=|\lambda_2^{m-2}|=.....=|\lambda_{n-1}^{m-2}|,...,$$

$$|\lambda_1^{m-i-1}|=|\lambda_2^{m-i-1}|=....=|\lambda_{n-i}^{m-i-1}|,...,$$

$$|\lambda_1^{m-n+1}|=|\lambda_2^{m-n+1}|. \quad (3.9a)$$

The system of equations (3.8), (3.9) defines the sought (*m*-1) moments of the controls discretization.
In a particular, from equation (3.8) the relation (3.8a) follows, which is inconsistent with the condition of a pair-wise equalization of the eigenvalues (3.3) at *n>m*.
The system (3.9) is a well defined, it agrees with (3.1),(3.2) and coincides with (3.8) *if* the number of its equations equals to the number of the equation state's variables. Thus, equations (3.7), (3.8), (3.9) have a sense only when *n=m*. The *n*-dimensional process requires *n* discrete controls applied at (*n-1*) intervals, defined by (3.8), (3.3) at the given starting conditions for equations (3.2). •

*Remark.* In the case of the matrix' renovation, each following solution (3.5) begins with a renovated eigenvalue, forming the chain (3.8), (3.9).

*Theorem* 3.2 (T3.2). The fulfillment of conditions (3.3) leads to an indistinctness in time of the corresponding *transformed* state's variables:

$$\hat{z}_i = \hat{z}_j, \begin{pmatrix} z_i \\ z_j \end{pmatrix} = \hat{G}_{ij} \begin{pmatrix} \hat{z}_i \\ \hat{z}_j \end{pmatrix}; \hat{G}_{ij} = \begin{pmatrix} \cos\varphi_{ij}, -\sin\varphi_{ij} \\ \sin\varphi_{ij}, \cos\varphi_{ij} \end{pmatrix},$$

$$\varphi_{ij} = arctg(\frac{z_j(\tau_k) - z_i(\tau_k)}{z_j(\tau_k) + z_i(\tau_k)}) \pm N\pi, N = 0,1,2... \quad (3.10)$$

in some coordinate system, built on the states $(0z_1...z_n)$ and rotated on angle $\varphi_{ij}$ in (3.10).

To *prove* we consider the geometrical meaning of the condition of equalizing of the eigenvalues as a result of the solutions of the equations (3.1), (3.2).
Applying relations (3.3) to the solutions of (3.8) for a nearest $i, j$, $i \neq j$, we get

$$\frac{dz_i}{z_i dt} = \frac{dz_j}{z_j dt} \; ; \; z_j(t,\bullet) = \frac{z_j(\tau_k.)}{z_i(\tau_k.)} z_i(t,\bullet), \; i,j=1,...,n, \; k=1,...,(n-1), \quad (3.10a)$$

where the last equality defines a hyper plane, being in a parallel to the axis $z_i = 0, z_j = 0$

in coordinate system $(0z_1...z_n)$. By rotating this coordinate system with respect to that axis, it is found a coordinate system where the equations (3.10a) are transformed into the equalities for the state variables $\hat{z}_i$ in form (3.10). The corresponding angle of rotation of coordinate plane $(0z_i z_j)$ is determined by relation (3.10).

Due to the arbitrariness of $k = 1,...,(n-1)$, $i, j = 1,...,n$ the foregoing holds true also for any two components of the state vector and for each interval of discretization. By carrying out the sequence of such $(n-1)$ rotations, we come to the system $(0\hat{z}_1...\hat{z}_n)$, where all the state variables are indistinguishable in time. •

*Comments.* If a set of the discrete moments $(\tau_k^1, \tau_k^i, \tau_k^{N_k})$ exists (for each optimal control $v_k$) then a unique solution of the optimization problem is reached by choosing a minimal interval $\tau_k^i$ for each $v_k$, which accomplishes the transformation of the above system to the origin of coordinate system during a minimal time. The macrovariables are derived as a result of memorizing of the states $z_i(\tau_k)$, $i, k = 1,...,n$, being an attribute of the applied control in (3.2), which are fixed along the extremal segments.

The transformation $(G \times \hat{G}_{ij})$ transfers $\{x_i\}$ to new macrovariables $\{\hat{z}_i\}$, whose pair-wise indistinctness at the successive moments $\{\tau_k\}$ agrees with the reduction of numbers of independent macrocoordinates. This reduction has been referred as the *states' consolidation*.

The successive equalization of the relative phase speed in (3.10a), accompanied by *memorization* of $z_i(\tau_k)$, determines an essence of the mechanism of the states' ordering [4,8].

Therefore, the problem of forming a sequentially *consolidated* macromodel is solved in a real–time process of the optimal motion, combined with identification of the renovated operator. Whereas both equalization and cooperation follow from the solution of the optimal problem for the path functional.

The macromodel is reversible within the discrete intervals and is irreversible out of them. Thus, a general structure of the initial object (1.1)(used also in physics), allows modeling a wide class of complex objects with *superimposing* processes, described by the *equations of irreversible thermodynamics* ([2], sec.7).

According to the extremal properties of the information entropy, the segments of the extremals approximate the stochastic process with a maximal probability, i.e., without losing information about it. This also allows us to get the *optimal and nonlinear filtration* of the stochastic process within the discrete intervals [4].

The model dynamics is initiated by applying a starting step-wise control in the form

$$v(\tau_o^o) = -2E_{\tau_o^o}[\tilde{x}_t(s)], \tag{3.11}$$

at $\tau_o^o = s + o$, where $\tau_o^o$ is the moment of the control's starts, $\tilde{x}_t(s)$ are the object's initial conditions, which also include given correlations $r(s) = E[\tilde{x}_t(s)\tilde{x}_t(s)^T]$ and/or $b(s) = 1/2\dot{r}(s)$.

These initial conditions also determine a starting *external* control

$$u(\tau_o^o) = b(\tau_o^o)r(\tau_o^o)^{-1}v(\tau_o^o), \tag{3.12}$$

where $v(\tau_o^o) = -2x(\tau_o^o)$, and a nonrandom state can be defined via

$$x(\tau_o^o) \cong |r^{1/2}(\tau_o^o)|. \tag{3.13}$$

This control imposes the constraint (2.18a) in the form (2.18d) that allows starting the dynamic process.
The above initial conditions identify

$$A(\tau_o^o) = b(\tau_o^o)r(\tau_o^o)^{-1}, A(\tau_o^o) = (\lambda_i(\tau_o^o)), i = 1,...,n \tag{3.14}$$

which is used to find a first time interval $t_1 = \tau_1^1 - \tau_o^o$ between the punched localities, where the next matrix's elements is identified, and so on.

*Specific* of the considered optimal process *consists of the computation of each following time interval* (where the identification of the object's operator will take place and the next optimal control is applied) *during the optimal movement under the current optimal control, formed by a simple function of dynamic states.* In this optimal dual strategy, the IPF optimum predicts each extremal's segments movement not only in terms of a total functional path goal, but also by setting at each following segment the renovated values of this functional's controllable shift and diffusion, identified during the optimal movement, which currently correct this goal.

## 4. The consolidation of the model's processes in a cooperative information network (IN). The IN code

Conventional information science, considering generally an information *process*, traditionally uses the probability measure for the random *states* and corresponding Shannon's entropy measure as the uncertainty's *function* of the states [21, 16,15, other].

The entropy *functional* defines the conditional quantity of information for the compared stochastic *processes* $\tilde{x}_t$, $\tilde{x}_t^1$, and the IPF allows building a *dynamic information network* for the corresponding *macroprocesses*.

The fulfillment of condition (2.26) connects the extremal segments of a multi-dimensional process leading to the segment's *cooperation*, while the realization of condition (3.10) reduces a number of the model's independent states carrying a state's *consolidation*. Both these specifics allow *grouping* the cooperative macroparameters and *aggregating* of their equivalent dimensions in an ordered *hierarchical information network* (IN), built on a multi-dimensional spectrum of the system's operator, which is identified during the optimal motion.

The IN organization includes the following steps: arranging the extremal segments in an ordered sequence; finding an optimal mechanism of connecting the arranged segments into a sequence of their consolidating states, whose joint dimensions would be sequentially deducted from the initial model's dimension; and forming an hierarchy of the adjoining cooperating dimensions. Below we consider the formal relations and procedure implementing these steps, which are based on the variation's and invariant conditions following from the initial VP (sec.2). We illustrate these relations using the *n*-dimensional spectrum of the complex eigenvalues $\lambda_{io} = \alpha_{io} \pm j\beta_{io}$ for the model's starting matrix $A(t_o)$, which we assume all different with the ratio $\gamma_{io} = \beta_{io}/\alpha_{io}, \alpha_{io} \neq 0$, $i = 1,.....,n$. The segments' cooperation produces a chain of the matrix's eigenvalues $A(t_k, t_k + o) = (\lambda_{it}^k, \lambda_{it}^k + o)$ with $\lambda_{it}^k = \alpha_{it}^k \pm j\beta_{it}^k$ and $(\lambda_{it}^k + o) = (\alpha_{it}^k + o) \pm j(\beta_{it}^k + o)$ at each segment's end and a beginning of a following segment accordingly; $k = 1,.....,N$ is the number of DP($t_k + o$) where cooperation of $\lambda_{it}^k$ and $\lambda_{it}^k + o$ takes place.

A feasible IN joins of the multiple nodes, while each its node collects a group of the equal eigenvalues gained in the cooperative process. The optimal condition (2.12, 2.18, 2.36) for such groups of the eigenvalues, considered at a moment of cooperation $t_k + o$, acquires the form

$$\min Tr[\lambda_{it}^k(t_k + o)] = \min[\sum_{r=1}^{m} g_r \lambda_r^g], \qquad (4.1)$$

where $g_r$ is a $r$-th group with it's a joint eigenvalue $\lambda_r^g$, $m$ is a total number of groups-the IN's nodes. Cooperation of the corresponding *states's variables* is carried by applying transformation (2.10) to the related $\lambda_r^g$. For realization of (4.1) we apply the invariant (2.34), which is concretized in a form

$$\lambda_{io}(t_{it} - t_{io}) = \lambda_{io}t_{it} - \lambda_{io}t_{io} = inv, \qquad (4.2)$$

where $\lambda_{io}$ is fixed at each moment $t_{io}$ of the segment's beginning leading to

$$\lambda_{io}t_{io} = inv \text{ and } \lambda_{io}t_{it} = inv_o. \qquad (4.2a)$$

By the $\Delta t_{io}$ interval's end $t_{it} = t_i$, the eigenvalue $\lambda_i(t_i) = \lambda_{it}$ satisfies the following solution at the applied control:

$$\lambda_i(t_i) = \lambda_{it} = \lambda_{io}\exp(\lambda_{io}\Delta t_{io})(2 - \exp(\lambda_{io}\Delta t_{io}))^{-1}, i = 1,..,k,...,n. \qquad (4.3)$$

Substituting invariants (4.2), (4.2a) in (4.3) we get $\lambda_{it}t_{it} = (\lambda_{io}t_{it})inv(\exp) = inv_o inv(\exp) = inv_1$. (4.3a)

Applying (4.2,4.2a) to the initial model's complex eigenvalues, we come to the local invariants

$$\alpha_{io}t_i = inv = \mathbf{a}_o, \beta_{io}t_i = inv = \mathbf{b}_o, \alpha_i t_i = inv = \mathbf{a}, \beta_i t_i = inv = \mathbf{b}, \ t_i = t_{it}, i = 1,...,n, \quad (4.4)$$

with $\alpha_{io}, \alpha_i$ and $\beta_{io}, \beta_i$ representing the real and imaginary information speeds (according to relations (2.14,2.14a)), while the invariants $\mathbf{a}_o, \mathbf{b}_o$ measure the quantity of real and imaginary information, produced *during* the interval by its end; invariants $\mathbf{a}, \mathbf{b}$ measure the quantity of real and imaginary information, produced *at the ending moment* the interval, prior of the segment's cooperation.

Both invariants are used constructively in building the cooperating chain of the eigenvalues satisfying the condition (2.26) in the forms:

$$|\lambda_i(t_k)| = |\lambda_j(t_k + o)|, \quad (4.5)$$

for each cooperating $i,k$ segments, whose eigenvalues satisfy the solutions (4.3).

A successive application of (4.3, 4.5) brings a cooperation of the extremals eigenvalues' segments at each DP. Solving jointly (4.3, 4.5) with the invariant relations (4.4), we find the eqs for the invariants at the DP:

$$2(\sin(\gamma \mathbf{a}_o) + \gamma \cos(\gamma \mathbf{a}_o)) - \gamma \exp(\mathbf{a}_o) = 0, \ \gamma_{io} = \frac{\beta_{io}}{\alpha_{io}}, \ \alpha_{io} \neq 0 \ \beta_{io} \neq 0, \ \text{at} \operatorname{Im}\lambda_j(t_k) = 0, \quad (4.6)$$

Using relation $|\lambda_i^t(t_i)t_i|^2 = \mathbf{a}^2$ and the representation (4.3) via the invariants, we get these invariants' connection by

$$\mathbf{a} = \mathbf{a}_o \exp(-\mathbf{a}_o)(1-\gamma^2)^{1/2}(4 - 4\exp(-\mathbf{a}_o)\cos(\gamma \mathbf{a}_o) + \exp(-2\mathbf{a}_o))^{-1/2}. \quad (4.7)$$

This allows us to evaluate both invariants. From the solution of (4.6) at $\gamma \to 1$ we get a minimal $\mathbf{a}_o(\gamma = 1) \to 0$, which brings also the minimal $\mathbf{a}(\gamma = 1) = 0$ from (4.7). The first one at $\gamma \to 0$ limits a maximal quantity of a real information produced at each segment; the second one at $\gamma \to 1$ restricts to a minimum the information contribution necessary for cooperation and, therefore, puts a limit on the information cooperation. It's also seen that relation (4.6) as the function of $\gamma$ reaches its extreme at $\gamma = 0$, which at $|\mathbf{a}_o(\gamma = 0)| = 0.768$, brings $\mathbf{a}(\gamma = 0) \cong 0.23193$.

Actually, a feasible admissible diapason of $\gamma_{io} = \gamma$, following from the model simulation [33], is $0.00718 \leq \gamma_{io} \leq 0.8$ with the condition of a model equilibrium at $\gamma = 0.5$, $\mathbf{a}_o(\gamma = 0.5) \cong \ln 2$, $\mathbf{a}(\gamma = 0.5) \cong 0.25$.

The cooperation of the real eigenvalues, according to (4.7b), reduces the condition (4.1) to the form

$$\min[\sum_{r=1}^R g_r \lambda_r^g] = \min[\sum_{r=1}^R g_r \alpha_r^g], \quad (4.8)$$

where $\alpha_r^g > 0$ is a joint real eigenvalue for each group, satisfying the requirement of positive eigenvalue $\alpha_r^g$ at applying the optimal control (2.29,2.30). A number of the joint eigenvalues in a group $g_r$ we find starting with a doublet as a minimal cooperative unit (fig.1). The minimal $\alpha_r^g$ for the doublet with two starting real eigenvalues at $|\alpha_{io}| < |\alpha_{ko}|$ can be reached, if by the moment $t_i$ when at $\alpha_{it} = \mathbf{a}_o / t_i$, the initial eigenvalue $\alpha_{ko}$ brings the eq. (4.4) to the form $\alpha_k(t_i) = \alpha_{ko}\exp(\alpha_{ko}(t_i - t_{ko}))[2 - \exp(\alpha_{ko}(t_i - t_{ko}))]^{-1}, \alpha_{ko} = \alpha_{ko}(t_{ko})$, whose solution $\alpha_k(t_i)$ will coincide with $\alpha_{it}$ by the end of the $t_i$ duration, and $\alpha_r^g = 2\alpha_{it}$.

The fulfillment of $|\alpha_{it}|/|\alpha_{io}| = |\mathbf{a}/\mathbf{a}_o|$ at $\gamma = (0.0 - 0.8)$, $|\mathbf{a}| < |\mathbf{a}_o|$ guarantees the decreasing of both $\alpha_{it}$ and $\alpha_k(t_i)$, fulfilling the inequalities

$$|\alpha_{it}| < |\alpha_{io}|, \ |\alpha_k(t_i)| < |\alpha_{ko}|. \quad (4.9)$$

Let's consider also a triplet as an elementary group of the cooperating three segments with the initial eigenvalues $|\alpha_{jo}|<|\alpha_{io}|<|\alpha_{ko}|$, where the minimal eigenvalue $\alpha_{jo}$ of a third segment we add to the previous doublet (for a convenience of the comparison)(fig.1). Then the minimal $\alpha_r^g$ can be reached (at other equal conditions) if the moment for joining of the first two eigenvalues (with the initials $|\alpha_{io}|<|\alpha_{ko}|$) coincides with the moment $t_j$ of forming the minimal $\alpha_{jt} = |\mathbf{a}_o|/t_j$ for the third eigenvalue.

Then the additional discrete interval is not required.

Compared with the related doublet, we have $|\alpha_{jt}|<|\alpha_{it}|$, where each minimal eigenvalue is limited by a given ranged initial spectrum. Therefore, a minimal optimal cooperative group is a triplet with $\alpha_r^g = 3\alpha_{jt}$.

For a space distributed macromodel [8], a number of cooperating segments is three, each one for every space dimension. This limits also a maximal number of the cooperating segments by three in each dimension for every elementary cooperative unit. The selection of the triplet's sequence and their arrangement into the IN is possible after ranging the initial $(\alpha_{io})_{i=1}^n$ in their decreasing values:

$|\alpha_{1o}|>|\alpha_{2o}|>,....|\alpha_{io}|>|\alpha_{i+1o}|,......>|\alpha_{no}|$. (4.9a)

Applying the needle controls at the moment of cooperation (for example at $(t_i + o)$ for the doublet) takes place when, in addition to the execution of (4.4) in the form $|\alpha_i(t_i + o)|=|\alpha_k(t_i + o)|$, and the reaching a minimum among the sum of the egenvalues prior the cooperation:
$|\alpha_i(t_i + o)|+|\alpha_k(t_i + o)|= 2|\alpha_i(t_i + o)|=|\alpha_i^g|= \min(|\alpha_i(t_i)|+|\alpha_k(t_k)|)$, (4.10)

the cooperated sum satisfies also a maximum condition regarding any sum of the following two eigenvalues:
$2|\alpha_k(t_i + o)|=|\alpha_i^g|= \max[|\alpha_{i+1}(t_{i+1})|+|\alpha_{i+2}(t_{i+2})|]$. (4.11)

Because for the ranged $(\alpha_{io})_{i=1}^n$ conditions (4.10,4.11) is satisfied, there are also fulfilled the relations $(|\alpha_{i-1,o}|+|\alpha_{io}|) = \max[|\alpha_{i+1,o}|+|\alpha_{i+2,o}|]$, as well as $|\alpha_{io}|= \max[|\alpha_{i+1o}|]$.

The formalization of this procedure leads to a minimax representation of eigenvalues by the Kurant-Fisher's variation theorem [22], which brings the condition of a sequential ranging for the macromodel eiegenvalues' spectrum. The result follows from a successive application of the maximum condition to the minimal condition for the Relay's ratio $q(x) = \dfrac{(x, Ax)}{(x, x)}$ for a macromodel's matrix $A > 0$, which leads to

$|\lambda_i| \geq \dfrac{(x, A_i x)}{(x, x)} \geq |\lambda_{i+1}|,....$, or in our case to $|\alpha_{i-1}^g|>|\alpha_i^g|>|\alpha_{i+1}^g|$ . The geometrical meaning illustrates an ellipsoid, whose axes represent the model's eigenvalues. The method starts with the maximal eigenvalue $|\alpha_1(t_1)|= \max_x q(x)$, taken from a maximal axis of the ellipsoid's orthogonal cross section, that is rotating up to reaching a minimal axis of the ellipsoid's cross section, which should coincide with the following lesser eigenvalue $|\alpha_2(t_2)|<|\alpha_1(t_1)|$, and so on. Because the model works with ranging the current $|\alpha_i(t_i)|$, the procedure brings also a monotonous sequence of the starting eigenvalues in (4.9a).

There are two options in forming the IN:

(1)-identify the IN by collecting the current number of equal $\alpha_r^g$ for each cooperative group $g_r$-as an IN node, and then arranging these nodes into a whole IN;

(2)-building an optimal IN by collecting the triplet's $\alpha_r^{g=3} = \alpha_r^3$ and using invariant relations (4.4,4.6a, 4.7).

In both cases, the current eigenvalues are identified under the action of the applied control by relations $\dot{r}_i r_i^{-1} = 2\lambda_i$, where $r_i(t) = E[\tilde{x}_i(t)^2]$, $i = 1,...,n$ are the covariation functions, and $\tilde{x}_i(t)$ is the observed microlevel's process. These eigenvalues allow us also determine the invariants (4.4), calculate $\alpha_r^3$, and the triplet's number for the optimal IN with $n$ initial eigenvalues.

The sequential cooperation of the ranged eigenvalues by threes, leads to the repeating of the initial triplet's cooperative process for each following triplet with the preservation of two basic eigenvalues ratios $\gamma_1^\alpha = \frac{\alpha_{1o}}{\alpha_{2o}} \to \frac{\alpha_{io}}{\alpha_{i+1o}}$, $\gamma_2^\alpha = \frac{\alpha_{2o}}{\alpha_{3o}} \to \frac{\alpha_{i+1,o}}{\alpha_{i+2,o}}$ satisfying the equations

$$\gamma_1^\alpha = \frac{\exp(\mathbf{a}(\gamma)\gamma_1^\alpha\gamma_2^\alpha) - 0.5\exp(\mathbf{a}(\gamma))}{\exp(\mathbf{a}(\gamma)\gamma_2^\alpha) - 0.5\exp(\mathbf{a}(\gamma))}, \gamma_2^\alpha = 1 + \frac{\gamma_1^\alpha - 1}{\gamma_1^\alpha - 2\mathbf{a}(\gamma)(\gamma_1^\alpha - 1)}, \quad (4.12)$$

where parameter $\gamma$ is found from relation (4.7) via known the invariants (4.4), which are the common for the optimal model as well as the $\gamma$ is.

The system of equations (4.3-4.12) allows the restoration of the model's macrodynamics $x_i(t) = F(x_{io}, t, \tau_i, \lambda_i^t(t))$ by knowing the initial $\lambda_{io}$, $x_{io}$, and finding $\tau_i$ via the invariants, which also determine $\gamma_1^\alpha$, $\gamma$ and as a result the structure of optimal IN.

The implementation of the above equations leads to a creation of the successively integrated information macrostructures that accompany the increasing of the intervals' sequence $\tau_i$, $i = 1,...,N$ and decreasing of the consolidated real eigenvalue $\alpha_r^3 = \mathbf{a}_o/t_r$, $r = 3,5,7,...n$. The sequence of the cooperating ordered eigenvalues $\alpha_r^3$, $r = 3,5,7,..m$ moves to its minimal $\alpha_m^3$ with the IN minimal dimension for a final node $n_{mo} \to 1$. The optimal IN's triplet structure includes the doublet as a *primary* element with adding a third eigenvalue to the first doublet, and then adding to each consolidated $\alpha_r^3$ the following doublet. The considered sequence of the triplet's optimal processes, transfers the consolidating $\alpha_r^3$ on the switching control line ($\mathbf{a} = \alpha_m^3 t_m = inv$), at which the minimal $|\alpha_m^3|$ for each $m$-node will be achieved at the node's cooperative moment $t_m$. This strategy is executed for the spectrum of the initial eigenvalues, defined by the multiplicators (4.12) with the maximal

$$\gamma_n^{\alpha_0} = (2.21)(3.89)^{n/2}, \gamma = 0.5. \quad (4.13)$$

Within the distance between the eigenvalues' spectrum, determined by above $\gamma_1^\alpha, \gamma_2^\alpha$, the model represents an *optimal-minimal filter*. The values $(\alpha_{io})$ that are different from the optimal set

$$\alpha_{i+1,o} = (0.2567)^i (0.4514)^{1-i} \alpha_{1o}, \alpha_{1o} = \alpha_{\max}, \gamma = 0.5 \quad (4.14)$$

are filtered and do not affect the IN peculiarities in the practical implementation. At known $\alpha_{1o}$, and a given $(n, \gamma)$, can be found the spectrum of the initial eigenvalues including $\alpha_{no}$, the invariants, $\gamma_1^\alpha, \gamma_2^\alpha$, and the IN's structure is build for the optimal macromodel without using the microlevel [8].

***The triplet's genetic code.*** The model possesses two scales of time: a reversible time that equals to the summary of the time intervals on the extremals $T^r = \sum_{i=1}^{i=n} t_i$, and the irreversible life-time $T_e^{ir}$ that is counted by the summary of the irreversible time intervals $\delta(t_i)$ between the extremals window's reversible time intervals.

*Proposition* 4.1. The ratio of the above elementary time intervals is evaluated by formula

$$-\frac{\delta t_i}{t_i} = (\frac{\Delta S_i^\delta}{\mathbf{a}_o^2} - 1), \quad (4.15)$$

where $\Delta S_i^\delta$ is an information contribution delivered during the reversible time interval $t_i$, $\mathbf{a}_o$-invariant.

*Proof.* Because the needle control connects the extremal segments by transferring information between the segment's window $\delta(t_i)$, i.e. from the $i$-segment's information $\Delta S_i^\delta / t_i$ to the $(i+o)$-segment's information $\Delta S_i^\delta /(t_i + \delta t_i)$, the information contribution from the needle control $\delta\alpha_i^o$, delivered during $\delta(t_i)$ is

$$\Delta S_i^\delta / t_i - \Delta S_i^\delta /(t_i + \delta t_i) = \frac{\Delta S_i^\delta \delta t_i}{t_i^2 + t_i \delta t_i} = \delta\alpha_i^o. \tag{4.16}$$

From other consideration, $\delta\alpha_i^o$ is evaluated by an increment of information production at $\delta(t_i)$:

$$\delta\alpha_i^o = \delta\frac{\partial \Delta S_{io}^\delta}{\partial t} \simeq \frac{\partial^2 \Delta S_{io}^\delta}{\partial t^2} \delta t_i, \text{ where}$$

$$\frac{\partial^2 \Delta S_{io}^\delta}{\partial t^2} = \lim_{\delta t_i \to 0} \frac{\partial^2 \Delta S_i}{\partial t^2}(\delta t_i) = -\frac{\partial H_{io}}{\partial t}, H_{io} = 1/2\alpha_i^t, \dot{\alpha}_i^t = -2(\alpha_{i-1}^t)^2 \exp(\alpha_{i-1}^t t_i)(2 - \exp(\alpha_{i-1}^t t_i))^{-2}, \tag{4.17}$$

$$\lim_{t_i \to 0} \dot{\alpha}_i^t = -2(\alpha_{i-1}^t)^2, \dot{H}_{io} = -(\alpha_{i-1}^t)^2, \delta\alpha_i^o = (\alpha_{i-1}^t)^2 \delta t. \tag{4.17a}$$

By substituting (4.17) into (4.16), at $\mathbf{a}_o(\gamma) = \alpha_{i-1}^t t_i$, we get (4.15). •

<u>Proposition</u> 4.2 (1)-Each extremal segment's interval $t_i$ retains $\mathbf{a}_o(\gamma)$ units of the information entropy;

(2)-The regular control brings negentropy $\mathbf{a}(\gamma) = \alpha_i^t t_i$ for interval $t_i$, where the control is memorized at each of the segment's locality.

*Proofs* of (1),(2) follow from the invariant relations (4.4) and an essence of the control's actions. •

<u>Corollary</u> 4.1. By evaluating the information contribution on the $t_i$-extremal by both the segment entropy's invariant $\mathbf{a}_o$ and the regular control's negentropy invariant $\mathbf{a}$, we come to

$$\Delta S_i^\delta = \mathbf{a}_o - \mathbf{a}, \text{ and } \frac{\delta t_i}{t_i} = \frac{\mathbf{a}_o - \mathbf{a} - \mathbf{a}_o^2}{\mathbf{a}_o^2} = \delta^*(\gamma) \bullet \tag{4.17b}$$

<u>Corollary</u> 4.2. The model's life-time ratio $T_*^{ir} = T^{ir}/T^r$ is evaluated by the invariant ratio $\delta^*(\gamma)$ at $t_i$-extremal:

$$T_*^{ir} = \frac{\mathbf{a}_o - \mathbf{a} - \mathbf{a}_o^2}{\mathbf{a}_o^2} = \delta^*(\gamma) \tag{4.18}$$

*Indeed.* Using $\delta t_i = \frac{\mathbf{a}_o - \mathbf{a} - \mathbf{a}_o^2}{\mathbf{a}_o^2} t_i$, $T^{ir} = \sum_{i=1}^n \delta t_i$ and $T^r = \sum_{i=1}^{i=n} t_i$, we come to $T_*^{ir} = \frac{\mathbf{a}_o - \mathbf{a} - \mathbf{a}_o^2}{\mathbf{a}_o^2}$. •

<u>Comments</u> 4.1. Let's count $\delta^*(\gamma)$ at $\gamma \in (0, 1)$ and $T^r = \sum_{i=1}^{i=n} t_i \simeq 2 t_{n-1}$. Then the $\delta^*(\gamma)$-function takes the values from 0.0908, at $\gamma = 0.1$, to 0.848, at $\gamma = 1$, with a minimal value 0.089179639, at $\gamma = 0.5$.

At $n=22$, $t_{n-1}=2642$, we have $T_e^{ir} = 471.225$, at $\gamma = 0.5$, with a maximal $T_e^{ir} = 4480.832$. •

<u>Corollary</u> 4.3. A minimal $\frac{\delta t_i}{t_i} \to 0$ leads to equality

$$\mathbf{a}_o(\gamma) - \mathbf{a}(\gamma) - \mathbf{a}_o^2(\gamma) \simeq 0, \tag{4.19}$$

which is approximated with an accuracy $\delta^* \mathbf{a}_o^2 = 0.044465455$, $(\gamma = 0.5)$. • \hfill (4.19a)

<u>Corollary</u> 4.4. Because each extremal segment's $t_i$ interval retains $\mathbf{a}_o(\gamma)$ units of the information entropy, and the regular control brings negentropy $\mathbf{a}(\gamma) = \alpha_i^t t_i$ for the interval $t_i$, while a needle control is also applied on this interval, the fulfillment of eq (4.19) means that the information contribution, delivered by needle control for interval $t_i$, is evaluated by invariant $\mathbf{a}_o^2(\gamma)$. Therefore (4.19) expresses the balance of information at each $t_i$-extremal at condition $\frac{\delta t_i}{t_i} \to 0$, which at $\gamma = 0.5$ is approximated with the accuracy (4.19a). •

*Comments* 4.2. Since the needle control joins the extremal segments by delivering information $\mathbf{a}_o^2(\gamma)$, we might assume that $\delta^* \mathbf{a}_o^2$ represents a defect of the $\mathbf{a}_o^2$ information, which is sealed after cooperation. Taking this into account, leads to a precise fulfillment of the balance equation in the form
$$\mathbf{a}_o(\gamma) - \mathbf{a}(\gamma) - \mathbf{a}_o^2(\gamma) - \delta^* \mathbf{a}_o^2(\gamma) = 0, \tag{4.19b}$$
while $\delta^* \mathbf{a}_o^2$ encloses the information spent on the segments' cooperation. •

*Proposition* 4.3. *The information structure of a triplet.*
A triplet, formed by the three-segments cooperative dynamics during a *minimal* time, encloses information $4\mathbf{a}_o^2 + 3\mathbf{a} \simeq 4$ bits at $\gamma = 0.5$, while each of the IN's triplet's *node* conceals information $\mathbf{a}_o^2 + \mathbf{a} \simeq 1$ bit.

*Proof.* The triplet's dynamics include two extremals, joining first into a doublet, which then cooperates with a third extremal segment (fig.1a). Forming the triplet during a minimal time requires building the doublet during the time interval of a third extremal segment, while all three dynamic processes start simultaneously with applying three starting controls. Each two extremals consist of two discrete intervals $(t_{i11}, t_{i12}, t_{i21}, t_{i22,})$ where $i$ is the triplet's number, $t_{i11}, t_{i12}$ are the first and second discrete intervals of the first dynamic process, $t_{i21}, t_{i22}$ are the first and second discrete intervals of the second dynamic process, $t_{i3}$ is a single discrete interval of the third dynamic process. The above requirement for a triplet with a minimal process' time implements the following equations on the first discrete interval for the first and second dynamics:
$$-\alpha_{i1o}t_{i11} + \alpha_{i11}t_{i11} + \mathbf{a}_o^2 = -\mathbf{a}_o + \mathbf{a} + \mathbf{a}_o^2 - \delta^* \mathbf{a}_o^2, \quad -\alpha_{i20}t_{i21} + \alpha_{i21}t_{i21} + \mathbf{a}_o^2 - \mathbf{a}_o + \mathbf{a} + \mathbf{a}_o^2 - \delta^* \mathbf{a}_o^2, \tag{4.20a}$$
where $-\alpha_{i1o}t_{i11} = -\mathbf{a}_o$, $-\alpha_{i20}t_{i21} = -\mathbf{a}_o$, $\alpha_{i21}t_{i21} = \mathbf{a}$, $\alpha_{i12}t_{i12} = \mathbf{a}$. This means that at each of these discrete intervals, the information balance is fulfilled with the accuracy $\delta^* \mathbf{a}_o^2$. The first and second dynamics, at the second time interval, convey the summary contribution $\alpha_{i13}t_{i13} + \alpha_{i23}t_{i23} + 2\delta^* \mathbf{a}_o^2$, followed by applying the needle control, which joins both dynamics into the doublet. This brings a balance condition in the form $\alpha_{i13}t_{i13} + \alpha_{i23}t_{i23} + 2\delta^* \mathbf{a}_o^2 = \mathbf{a}_o^2$. (We count here the information contribution from a defect $2\delta^* \mathbf{a}_o^2 = \delta^*(\gamma)$ at both intervals $t_{i21}, t_{i22}$). Joining the third segment's discrete interval with the doublet at the IN node requires applying another needle control, acting at the end of third interval (fig.1a).
This leads to the balance equation for third discrete interval in the form
$$-\alpha_{i30}t_{i31} + \alpha_{i31}t_{i31} + \mathbf{a}_o^2 - \delta^* \mathbf{a}_o^2 \simeq -\mathbf{a}_o + \mathbf{a} + \mathbf{a}_o^2 \simeq 0, \text{ at } \gamma = 0.5. \tag{4.20b}$$
It seen that a total information, delivered to the triplet, is equal $4\mathbf{a}_o^2 + 3\mathbf{a}$, which compensates for the information being spent on the triplet's cooperative dynamics:
$$3\mathbf{a}_o + \alpha_{i12}t_{i12} + \alpha_{i22}t_{i22} + \alpha_{i31}t_{i31} + \alpha_{i13}t_{i13} + \alpha_{i23}t_{i23} + 2\delta^* \mathbf{a}_o^2. \tag{4.20c}$$
Let's verify this result by a direct computation of the contributions $\alpha_{i13}t_{i13}$ and $\alpha_{i23}t_{i23}$, using the following formulas for each of them:
$$\alpha_{i13}t_{i13} = \alpha_{i12}(t_{i3} - t_{i11})\exp(\alpha_{i12}(t_{i3} - t_{i11}))[2 - \exp\alpha_{i12}(t_{i3} - t_{i11})]^{-1}$$
$$\alpha_{i23}t_{i23} = \alpha_{i12}(t_{i3} - t_{i21})\exp(\alpha_{i12}(t_{i3} - t_{i21}))[2 - \exp(\alpha_{i12}(t_{i3} - t_{i21})]^{-1}, \text{ where}$$
$$\alpha_{i12}(t_{i3} - t_{i11}) = \alpha_{i12}t_{i11}(t_{i3}/t_{i11} - 1) = \mathbf{a}(\gamma_{13} - 1), \ \alpha_{i22}(t_{i3} - t_{i21}) = \alpha_{i22}t_{i21}(t_{i3}/t_{i21} - 1) = \mathbf{a}(\gamma_{23} - 1), \tag{4.20d}$$

whose parameters at $\gamma=0.5$ takes the values $\gamma_{13}\simeq 3.9$, $\gamma_{23}\simeq 2.215$, $\mathbf{a}\simeq 0.252$.

The computation shows that the reqular controls, acting at $t_{13}$ and $t_{23}$, deliver information $\mathbf{a}(\gamma_{13}-1)=0.7708$ and $\mathbf{a}(\gamma_{23}-1)=0.306$ accordingly, while the macrodynamic process at these intervals consumes $\alpha_{i13}t_{i13}\simeq 0.232$ and $\alpha_{i23}t_{i23}\simeq 0.1797$. Including defect $\delta^*(\gamma)$, we get the information difference $\simeq 0.50088\simeq \mathbf{a}_o^2$ (at $\gamma=0.5$).

This means that both regular controls, acting on the second doublet's intervals, provide necessary information to produce the needle control, and therefore, the doublet satisfies the balance equation that does not need additional external information for the cooperation. The doublet's cooperation with the third extremal segment forms the triplet's IN node, which encloses the information contribution from both the doublet's and the third segment's needle controls (4.20b), providing the defect $\delta^*\mathbf{a}_o^2$ that satisfies the balance in (4.20b).

The triplet's information at $4\mathbf{a}_o^2+3\mathbf{a}\simeq 2.75$ (at $\gamma=0.5$) is measured in Nats (according to the basic formula for entropy[1]), or by $3.96\simeq 4$ bits in terms of $\log_2$ measure. Because the IN's triplet node consists of the doublet, which seals information $\delta^*\mathbf{a}_o^2$, and the third segment that transfers information $\mathbf{a}_o^2+\mathbf{a}-\delta^*\mathbf{a}_o^2\simeq 0.70535$ Nats to the node, the total node's information is $\mathbf{a}_o^2(\gamma)+\mathbf{a}(\gamma)\simeq 1.0157\simeq 1$ bits. •

*Comments 4.3.* The triplet's both regular and needle controls produce four switches (fig. 1a), which carry information $\simeq 4$ bits. Since each switch can encode one bit's symbol, or a single letter, it follows that a triplet is a carrier of a four letter's information code. This is a triplet's *genetic* code, initiated at the triplet's formation. Therefore, the generation of an *external* code with the same information switches, applied to the given initial eigenvalues spectrum, would be able to restore the triplet's information structure.

This means that such a code might reproduce the triplet's dynamics, which the genetic code had encoded. •

*Comments* 4.4. The triplet's information structure could serve as an information model of a DNA's triplet, which is the carrier of the DNA four letter's code. •

Finally, a time-space sequence of the applied controls, generating the doublets and triplets, represents a discrete control *logic*, which creates the IN's *code* as a *virtual communication language* and an algorithm of *minimal* program. The code is formed by a time-space sequence of the applied inner controls (both regular and needle), which automatically fixate the ranged discrete moments (DP) of the successive equalization of the model's phase speeds (eigenvalues) in the process of generation of the macromodel's spectrum. The equalized speeds and the corresponding macrostate's variables are *memorized* at the DPs by the optimal control actions. A selected sequence of the minimal nonrepeating DPs, produced by a *minimal ranged spectrum* of eigenvalues and the corresponding initial macrostates, generates the *optimal* IN's code, which initiates the ranged sequence of the IN's doublets, cooperating sequentially into triplets. The optimal code consists of a sequence of double states, memorized at the $\{t_i\}_{i=1}^{n-1}$ DP, and each $n$ dimensional model is encoded by **n** intervals of DP.

*Each triplet joins and memorizes three such discrete intervals, generating the triplet's digital code*

The above results reveal the procedure of the *transformation of a dynamic information into a triplet's code*. •

Cooperative information processes (macro associations) produce *information geometrical structures* [6], *distributed* in space (like various forms of pictures, images, etc.), which express their information contents *independently* on physical medium that materializes this information. This information geometry generally takes shapes of Riemann geometry, where an information process proceeds along a *geodesic line* on a particular surface, and a vector-speed depends on the surface's *curvature* [7].

An *information attraction* arises between these information geometrical structures, whose *intensity* is defined by the Riemann curvatures of the interacting information structures' geometries. Thus, the *space curvature* is both a result of *cooperative macrodynamics* and a *measure* of the information attraction, which, potentially, is physically materialized in gravitation. Moreover, a such *information surface* consists of a *cellular geometry*, where each cell encloses a code symbol, and whole surface's structure enfolds its a *genetic* code, which could be transmitted to other information structures during a mutual attraction and communication. (Such a cell is an elementary information model of a graviton) [7].

Any naturally made curvature conceals genetic information, which is a source of the curvature formation.

## 5. The IPF macromodel's singular points and the singular trajectories

For the considered concentrated and distributed macromodels, the analysis of the existence and uniqueness of singular points and/or the singular trajectories represents the principal and sufficient interest.

The pair equalization of the relative phase speeds at each discrete point for the concentrated model (2.26) leads to singularities of the dynamic operator, that follow from the consideration below.

We will show that the dynamics and geometry at each of the singular points of the spatial macromodel [8] are bound, and the singularities are associated with the model's cooperative phenomena.

Such analysis we provide for the model in partial derivations of the first order:

$$A_1 \frac{\partial x}{\partial t} = B_1 \frac{\partial x}{\partial l} + v_1; A_2 \frac{\partial y}{\partial t} = B_2 \frac{\partial y}{\partial l} + v_2 \qquad (5.1)$$

with variables $(x, y)$, spatial coordinate $l$, time $t$, controls $v_1(t,l,x)$, $v_2(t,l,y)$, and coefficients

$A_1 = A_1(t,l,x)$, $B_1 = B_1(t,l,x)$, $A_2 = A_2(t,l,y)$, $B_2 = B_2(t,l,y)$.

The initial conditions are given by the following distributions

$$x|_{t=s} = \varphi_1^l(l) = \tilde{x}(s,l_o,\tau,l), \text{ or } x|_{l=l_o} = \varphi_1^t(t); y|_{t=s} = \varphi_2^l(l) = \tilde{y}(s,l_o,\tau,l), \text{ or } y|_{l=l_o} = \varphi_2^t(t). \qquad (5.1a)$$

The equations (5.1), (5.1a) characterize a reflection of some region of plane $(t,l)$ on a region of space $(\Delta S, x, y)$, where the peculiarities and class of the surface $\Delta S = \Delta S(x,y)$ are completely defined by a specific equation of the reflection. At the known solution of problem (5.1, 5.1a), this surface's equation can be defined in a parametrical form:

$$x = x(t,l), y = y(t,l), \Delta S = \Delta S[x(t,l), y(t,l)]. \qquad (5.2)$$

For the given system, a singular point of the second order of the considered surface is determined by the condition of decreasing the following matrix's rank:

$$rank \begin{vmatrix} \frac{\partial x}{\partial t}, \frac{\partial y}{\partial t}, \frac{\partial \Delta S}{\partial t} \\ \frac{\partial x}{\partial l}, \frac{\partial y}{\partial l}, \frac{\partial \Delta S}{\partial l} \end{vmatrix} \neq 2, \qquad (5.3)$$

which corresponds to the identical turning to zero all minors of the second order of the above matrix.

By introducing the radius-vector $\vec{r} = x\vec{e}_1 + y\vec{e}_2 + \Delta S \vec{e}_3$ with the orths of basic vectors $\{\vec{e}_i\}_i^3$ and the derivatives $\vec{r}_t = \frac{\partial \vec{r}}{\partial t}, \vec{r}_l = \frac{\partial \vec{r}}{\partial l}$, we write (5.3) in the form

$$[\vec{r}_t \times \vec{r}_l] = 0. \tag{5.3a}$$

The equation of a normal $\vec{N}$ to surface (5.2) has view:

$$\vec{N} = \begin{vmatrix} \vec{e}_1, \vec{e}_2, \vec{e}_3 \\ \frac{\partial x}{\partial t}, \frac{\partial y}{\partial t}, \frac{\partial \Delta S}{\partial t} \\ \frac{\partial x}{\partial l}, \frac{\partial y}{\partial l}, \frac{\partial \Delta S}{\partial l} \end{vmatrix}, \text{ with the orth of the normal } \vec{N} : \vec{n} = \frac{\vec{N}}{N}. \tag{5.4}$$

Because $\vec{r}_t, \vec{r}_l$ are the tangent vectors to the coordinate lines, the fulfillment of (5.2) or (5.3) is an equivalent of the condition of a *nonexistence of the normal* (5.4) at the given singular point.

Since a normal to a surface is determined independently on a method of the surface parameterization, we come to the following conditions of the existence of the singular point:

$$\vec{N} = M_1 \vec{e}_1 + M_2 \vec{e}_2 + M_2 \vec{e}_2 = 0, \tag{5.5}$$

or

$$M_1 = \det \begin{vmatrix} x_t, y_t \\ x_l, y_l \end{vmatrix} = 0, \ \Delta S_l = \frac{\partial \Delta S}{\partial l}, \ x_t = \frac{\partial x}{\partial t}, y_t = \frac{\partial y}{\partial t}, \ x_l = \frac{\partial x}{\partial l}, y_l = \frac{\partial y}{\partial l}; \tag{5.5a}$$

$$M_2 = \det \begin{vmatrix} x_t, \Delta S_t \\ x_l, \Delta S_l \end{vmatrix} = 0, \ \Delta S_t = \frac{\partial \Delta S}{\partial t}, \ \Delta S_l = \frac{\partial \Delta S}{\partial l}; \tag{5.5b}$$

$$M_3 = \det \begin{vmatrix} y_t, \Delta S_t \\ y_l, \Delta S_l \end{vmatrix} = 0. \tag{5.5c}$$

According to (5.2) we have

$$\frac{\partial \Delta S}{\partial t} = \frac{\partial \Delta S}{\partial x} \frac{\partial x}{\partial t} + \frac{\partial \Delta S}{\partial y} \frac{\partial y}{\partial t}; \frac{\partial \Delta S}{\partial l} = \frac{\partial \Delta S}{\partial x} \frac{\partial x}{\partial l} + \frac{\partial \Delta S}{\partial y} \frac{\partial y}{\partial l}. \tag{5.6}$$

That's why relations (5.5a-c) are fulfilled automatically if (5.5) holds true.
Indeed, using (5.6) for (5.5b), we get

$$\frac{\partial x}{\partial t}(\frac{\partial \Delta S}{\partial x}\frac{\partial x}{\partial l} + \frac{\partial \Delta S}{\partial y}\frac{\partial y}{\partial l}) - \frac{\partial x}{\partial l}(\frac{\partial \Delta S}{\partial x}\frac{\partial x}{\partial t} + \frac{\partial \Delta S}{\partial y}\frac{\partial y}{\partial t}) = \frac{\partial \Delta S}{\partial x}(\frac{\partial x}{\partial t}\frac{\partial y}{\partial l} - \frac{\partial x}{\partial l}\frac{\partial y}{\partial t}) = \frac{\partial \Delta S}{\partial x} J, \tag{5.7}$$

where Jakobian for this system

$$J = \frac{D(x,y)}{D(t,l)} = \det \begin{vmatrix} x_t, y_t \\ x_l, y_l \end{vmatrix} = M_1 = 0. \tag{5.7a}$$

This brings the strong connection of the system's geometrical coordinates $(l, \vec{e})$ with the dynamics of $(x, y)$. Therefore, at a chosen representation (5.2), the singular points correspond also the degeneracy of the Jacobean J, or the fulfillment of condition

$$\frac{\partial x}{\partial t}\frac{\partial y}{\partial l} = \frac{\partial y}{\partial t}\frac{\partial x}{\partial l}, \tag{5.7b}$$

which for the distributed model is an analog of the equalization of the relative phase speeds (2.26).

Indeed. The analog of the relation (5.2) for the related matrix of the concentrated system is

$$rank \begin{vmatrix} \frac{dx}{dt}, \frac{d\Delta S}{dt} \\ \frac{dy}{dt}, \frac{d\Delta S}{dt} \end{vmatrix} \neq 2, \text{ which leads to}$$

$$\det \begin{vmatrix} \frac{dx}{dt}, \frac{d\Delta S}{dt} \\ \frac{dy}{dt}, \frac{d\Delta S}{dt} \end{vmatrix} = 0, \text{ or } \frac{d\Delta S}{dt}(\frac{dx}{dt} - \frac{dy}{dt}) = 0, \frac{d\Delta S}{dt} \neq 0, \frac{dx}{dt} = \frac{dy}{dt}. \tag{5.7c}$$

The last relation at $x(\tau) \simeq x(\tau + o) \neq 0$ coincides with (2.26) at $t = \tau$.

Let's apply (5.7b) to the system (5.1), written in the diagonal form:

$$(\lambda_1^t)^{-1}\frac{\partial x}{\partial t} = (\lambda_1^l)^{-1}\frac{\partial x}{\partial l} + v_1(t,l); \ (\lambda_2^t)^{-1}\frac{\partial y}{\partial t} = (\lambda_2^l)^{-1}\frac{\partial y}{\partial l} + v_2(t,l), \tag{5.8}$$

where $(\lambda_1^t, \lambda_1^l), (\lambda_2^t, \lambda_2^l)$ are the corresponding eigenvalues. For the diagonalized equations, it is possible to build the system of the regular differential equations in a symmetric form, generally

$$\frac{dt}{(\lambda_i^t)^{-1}} = -\frac{dl}{(\lambda_i^l)^{-1}} = \frac{dx_i}{v_i}, \ i=1,....,n \tag{5.9}$$

with its common integrals

$$\Phi^i = \Phi^i(\phi_1^i, \phi_2^i) = 0, \tag{5.10}$$

where the first integrals:

$$\phi_1^i = \int \lambda_i^l(l)dl + \int \lambda_i^t(t)dt, \ \phi_2^i = x_i + \int \lambda_i^l(l)v_i dl \tag{5.10a}$$

are the solutions of (5.9). The concrete form of the common integral is defined by (5.1a) and (5.8):

$$\Phi = \phi_2 - \int \lambda^l(l)[f(\phi)]v[f(\phi)]\frac{\partial f}{\partial \phi}d\phi + \varphi^l[f(\phi)], \tag{5.11}$$

where $\phi = \phi_1 - \int \lambda^\tau(\tau)d\tau = \int \lambda^l(l)dl + \int \lambda^t(t)dt - \int \lambda^\tau(\tau)d\tau; \int \lambda^\tau(\tau)d\tau = \int \lambda^t(t)dt |_{t=\tau}$,

$$\lambda^t = (\lambda_1^t, \lambda_2^t), \lambda^l = (\lambda_1^l, \lambda_2^l), \ x = (x_1, x_2), \Phi = (\Phi^1, \Phi^2), \phi_1 = (\phi_1^1, \phi_1^2), \phi_2 = (\phi_2^1, \phi_2^2) \tag{5.11a}$$

and $f$ is the root of equation $f = l(\phi_1(\tau), \tau)$ solved for $l$ and a fixed $t = \tau$:

$$\int \lambda^l(l)dl = \overline{\phi}_1 - \int \lambda^\tau(\tau)d\tau, \ \overline{\phi}_1 = \phi_1 |_{t=\tau}. \tag{5.12}$$

A partial solution of (5.9, 5.11-5.12) acquires the form:

$$x = -\int \lambda^l(l)v dl + \int \lambda^l(l)[f(\phi)]v[f(\phi)]\frac{\partial f}{\partial \phi}d\phi + \varphi^l[f(\phi)]. \tag{5.13}$$

Then, the corresponding partial derivations have the view:

$$\frac{\partial x}{\partial t} = -\int \lambda_1^l \frac{\partial v_1}{\partial t} dl + \Phi_1 \lambda_1^t, \quad \frac{\partial y}{\partial t} = -\int \lambda_2^l \frac{\partial v_2}{\partial t} dl + \Phi_2 \lambda_2^t \tag{5.14}$$

$$\frac{\partial x}{\partial l} = -\lambda_1^l v_1 + \Phi_1 \lambda_1^l, \quad \frac{\partial y}{\partial l} = -\lambda_2^l v_2 + \Phi_2 \lambda_2^l. \tag{5.14a}$$

$$\Phi_1 = [\frac{\partial \varphi_1^l}{\partial f_1}(\phi_1) + \lambda_1^l v_1(\phi_1)]\frac{\partial f_1}{\partial \phi_1} \tag{5.15}$$

By imposing the condition (5.7b) on the systems (5.14-5.15), we come to equation

$$(\int \lambda_1^l \frac{\partial v_1}{\partial t} dl - \Phi_1 \lambda_1^t)(\lambda_2^l v_2 - \Phi_2 \lambda_2^t) = (\int \lambda_2^l \frac{\partial v_2}{\partial t} dl - \Phi_2 \lambda_2^t)(\lambda_1^l v_1 - \Phi_1 \lambda_1^t) \quad (5.16)$$

which is fulfilled at the following cases:

$$\lambda_1^l = 0, \text{ or } \lambda_2^l = 0, \quad (5.16a)$$

and at the different combinations of the following pairs of the relations:

$$v_1 = \Phi_1, \text{ or } v_2 = \Phi_2 ; \quad (5.16b)$$

$$(\int \lambda_1^l \frac{\partial v_1}{\partial t} dl - \Phi_1 \lambda_1^t) = 0, \text{ or } (\int \lambda_2^l \frac{\partial v_2}{\partial t} dl - \Phi_2 \lambda_2^t) = 0; \quad (5.16c)$$

where the last two ones are correct if $\lambda_1^t \neq 0, \lambda_1^l \neq 0$, \quad (5.16d)

or, in particular, at the fulfillment of any of these relations:

$$\lambda_1^t = 0, \Phi_1 = 0, \frac{\partial v_1}{\partial t} = 0, \quad (5.16e)$$

$$\lambda_2^t = 0, \Phi_2 = 0, \frac{\partial v_2}{\partial t} = 0. \quad (5.16f)$$

Finally, we come to the condition:

$$\frac{(\int \lambda_1^l \frac{\partial v_1}{\partial t} dl - \Phi_1 \lambda_1^t)}{\lambda_1^l (v_1 - \Phi_1)} = \frac{(\int \lambda_2^l \frac{\partial v_2}{\partial t} dl - \Phi_2 \lambda_2^t)}{\lambda_2^l (v_2 - \Phi_2)} = \mathrm{I} . \quad (5.17)$$

It means, that for the *n*-dimensional PDE model (5.8) could exist an invariant condition (5.17) on the solution of (5.14,5.15), which is not dependable on the indexes in (5.17), or I could take a constant value for some pair of the indexes. If omit the trivial conditions (5.16a-5.16f) and the invariant (5.17), then (5.16) leads to the following relations:

$$\frac{\partial x}{\partial l} = \frac{\partial x}{\partial l} = 0, \text{ and } \frac{\partial y}{\partial l} = \frac{\partial y}{\partial t} = 0, \text{ or } \frac{\partial x}{\partial l} = \frac{\partial y}{\partial t} = 0 \text{ and } (5.7a).$$

The conditions (5.16) define the different equations of the singular points, or the singuar trajectories, created by any of the separated processes $x(t,l)$, or $y(t,l)$, while (5.17) defines the singular trajectory, created by the process' interactions. In such singularities, the rank of extended matrix (5.3) decreases that declines the number of independent equations in a system; and a normal to a surface $\Delta S$ at a singular point does not exist. Because of the eqs.(2.16) and (5.7b) connections, these conditions of singularities are applied also to the considered in secs.2,3 concentrated models. Therefore, the singular points, defined by the conditions (5.16) and (5.17) do exist, and they are not singles. The geometrical locations of the singular points could be the isolated states of the system (5.1), as well as the singular trajectories.

The invariant I corresponds to the equalization of the local subsystems relative speeds (at the phase trajectories) at transferring via the singular curve, being an analog of the condition (2.16) for the concentrated model. At these points, relation (5.16b) gets the form

$$\Phi_1 = [\frac{\partial \varphi_1^l}{\partial f_1}(\phi_1) + \lambda_1^l v_1(\phi_1)] \frac{\partial f_1}{\partial \phi_1} = v_1, \quad (5.18)$$

and it is fulfilled along the singular trajectory, in particular, at

$$\lambda_1^l = \mathrm{const}, \phi_1 = \lambda_1^t t + \lambda_1^l l - \lambda_1^\tau \tau, \ f_1 = \phi_1 (\lambda_1^l)^{-1} = l + (\lambda_1^t / \lambda_1^l)(t - \tau), \ \frac{\partial (f_1)}{\partial \phi_1} = (\lambda_1^l)^{-1}, \quad (5.18a)$$

which is satisfied at

$$\frac{\partial \varphi_1^l}{\partial f_1} = \frac{\partial \varphi_1^l}{\partial l}|_{l=f_1} = \lambda_1^l(v_1(t,l) - v_1(t,f_1)). \tag{5.18b}$$

This condition binds the automatic fulfillment of (5.6b) (at the macrotrajectories' singlular points) with the initial distribution (5.1a) (depending on the model's microlevel). That's why relations (5.16b-f), (5.17) might be considered as the *limitations* imposed on the class of the model's random processes, for example, applicable for Markov fields. At a given random field, which not satisfies these limitations, the conditions (5.16b.c) could be fulfilled by choosing the corresponding controls. At $\lambda_1^1 = \text{var}$, in particular at $\lambda_1^1(v_1)$, a possibility of the Jacobean degeneracy, as it follows from (5.18), is also covered by relations (5.18b).

From that follows that the model's singular phenomena could be implemented by the controls.

Therefore, the singular points and trajectories carry out the additional information about connection of the micro- and macroprocesses, the model's geometry, dynamics, and control. Because relations (2.16), (5.7a) are the conditions *connecting* the extremal's segments at the *o*-window, the singularities are related also to the model's cooperative phenomena. The state consolidation at the singular points is possible.

The detail analysis of the singular points is provided in [4] for a two dimensional concentrated model, where is shown that before the consolidation, the model has a saddle singular point, and after the consolidation its singular point becomes an attractor. More generally, the equalization of the subsystem's eigen frequencies (connected to the eigenvalues) (in (2.16), (5.7)) is an indicator of arising *oscillations*, which, at the superposition of the *diffusion* at the o-window, are faded into an *attractor*. Actually, applying just a regular control (as a first part of the needle control) at the model's *o*-window transfers the dynamic trajectories at the macrolevel to the random trajectories at the microlevel, while both of them are unstable. Applying a second regular control (being a second part of the needle control) brings stability to both of them.

Generally the model undergoes a global bifurcation at the *o*-window between the segments, under the control actions and by transferring from kinetics to a diffusion and then from the diffusion to kinetics.

Indeed. At the extremal's ending moment we have
$$a^u(\tau - o) = b(\tau - o)r^{-1}(\tau - o)x(\tau - o), \tag{5.19}$$
where $b(\tau - o)r^{-1}(\tau - o) = D_x(\tau - o)$ are the diffusion component of stochastic equation, which are compensated by the kinetic part, delivered with the regular control.

The needle control, applied between the segments at the moments $(\tau, \tau + o)$, brings the increment
$$\delta a^u = -a^u(\tau) + a^u(\tau + o) = -\lambda(\tau)x(\tau) + \lambda(\tau + o)x(\tau + o), \lambda(\tau - o) < 0, sign\lambda(\tau) = -sign\lambda(\tau - o), \tag{5.20}$$
which at $x(\tau + o) \approx x(\tau), \lambda(\tau + o) \approx -\lambda(\tau)$, determines $\delta a^u = -2\lambda(\tau)x(\tau)$.

Thus the needle control *decreases* the initial duffusion part $D_x(\tau - o) = \lambda(\tau - o)$ according to relation $b(\tau + o)r^{-1}(\tau + o) = D_x(\tau + o) \approx D_x(\tau) - 2\lambda(\tau) \approx -\lambda(\tau)$, transferring the diffusion into kinetics.

This means that applying of the needle controls to a sequence of the extremal segments increases an influence of the kinetics on the model, decreasing the diffusion components.

## 6. The natural variation problem, singular trajectories, and the field's invariants for the IPF

The information functional (IPF) of the distributed model in the form:

$$\Delta S = \iint_{\bar{G}} L\, dldt, \quad (6.1),\quad L = \frac{\partial x}{\partial t} X + \frac{\partial x}{\partial l} X, \tag{6.1a}$$

is defined on the controlled processes $x = \{x_i\}$, which are determined by the solutions of Euler-Ostrogradsky's equations for this functional and the natural border conditions, connected with the initial distributions (5.1a). Using the eqs (2.33a) we will use expression for

$X = 1/2 hx, h = r^{-1}, r = E_{\tilde{x},s,l_o}[\tilde{x}\tilde{x}^T]$ (6.1b) in the Lagrangian (6.1a).

The problem consists of synthesis of a control law $v = v(t,l,\dot{x}_t,\dot{x}_l)$ that carries out the fulfillment of extremal principle for the functional $\Delta S$ at the natural border conditions. This problem, which is called the natural variation problem, we solve for the equations having the structure (5.1) at the Lagrangian L in form (6.1a). This problem is aimed at its formal connection to an appearance of a singular curve (sec.5).

Writing the functional's variation at a variant control's definition domain $\bar{G}$, according to [23], we have

$$\delta\Delta S = \delta\Delta S_1 + \delta\Delta S_2 = \iint_{\bar{G}} \{\sum_{i=1}^{2}[\frac{\partial L}{\partial x_i} - \frac{\partial}{\partial t}(\frac{\partial L}{\partial \dot{x}_{it}}) - \frac{\partial}{\partial l}(\frac{\partial L}{\partial \dot{x}_{il}})]\delta\bar{x}_i\}dldt +$$

$$\iint_{\bar{G}} \{\sum_{i=1}^{2}[\frac{\partial}{\partial t}(\frac{\partial L}{\partial \dot{x}_{it}}) + \frac{\partial}{\partial l}(\frac{\partial L}{\partial \dot{x}_{il}})]\delta\bar{x}_i + \frac{\partial}{\partial t}(L\delta t) + \frac{\partial}{\partial l}(L\delta l)\}dldt = 0;\quad \delta\bar{x}_i = \delta x_i + \sum_{j=1}^{2} \frac{\partial x_i}{\partial l_j}\delta l_j; l_1 = t, l_2 = l \tag{6.2}$$

Condition $\delta\Delta S_1 = 0$ is fulfilled by the execution of Euler-Ostrogradsky's equation [18]

$$\frac{\partial L}{\partial x_i} - \frac{\partial}{\partial t}(\frac{\partial L}{\partial \dot{x}_{it}}) - \frac{\partial}{\partial l}(\frac{\partial L}{\partial \dot{x}_{il}}) = 0, \tag{6.3}$$

which for (5.1) and (6.1a,b),(5.2) acquires the forms

$$\frac{\partial x_i}{\partial t} h_{ii} + \frac{\partial x_i}{\partial t} x_i \frac{\partial h_{ii}}{\partial x_i} + \frac{\partial x_i}{\partial l} h_{ii} + \frac{\partial x_i}{\partial l} x_i \frac{\partial h_{ii}}{\partial x_i} - \frac{\partial}{\partial t}(h_{ii} x_i) - \frac{\partial}{\partial l}(h_{ii} x_i) = 0;$$

$$\frac{\partial h_{ii}}{\partial x_i} \frac{\partial x_i}{\partial t} + \frac{\partial h_{ii}}{\partial x_i} \frac{\partial x_i}{\partial l} = \frac{\partial h_{ii}}{\partial t} + \frac{\partial h_{ii}}{\partial l} \quad (\forall x_i \neq 0, \forall (l,t) \in G). \tag{6.4}$$

We get the equation of extremals

$$\frac{\partial h_{ii}}{\partial t} + \frac{\partial h_{ii}}{\partial l} = 0; \frac{\partial r_{ii}}{\partial t} + \frac{\partial r_{ii}}{\partial l} = 0; \tag{6.5}$$

At the solutions of this equation holds true the relation

$$E_i[L_i] = E_i[h_{ii} x_i (\dot{x}_{it} + \dot{x}_{il})] \equiv 0. \tag{6.6}$$

The condition $\frac{\partial^2 L}{\partial x_i^2} \neq 0$ at an extremal determines the regular, or not the singular extremal's points. For the linear regarding $(x,\dot{x}_t,\dot{x}_l)$ form (6.1a) is fulfilled $\frac{\partial^2 L}{\partial x_i^2} = 0 \ \forall (l,t) \in G$, and the obtained extremals are a non regular. At these extremals, the differential equations (6.3) turn into the parametrical equations for the functions $h_{ii}$ (6.4, 6.5,6.6) determined via $x(t,l) = \{x_i(t,l)\}$ in (6.1a). Applying the differential equation with the control processes $\{x_i = x_i(t,l)\}$, the piece-wised controls $\{v_i\}$, and random initial conditions, let's find the control in (5.1) for which the solutions of equations (5.1) satisfy to (6.5). Using (6.5) as the initial condition for the control synthesis, we get

$$L_i = h_{ii} x_i (\frac{\lambda_i^t}{\lambda_i^l} \dot{x}_{il} + u_i^t + \dot{x}_{il})]; \ M_i[L_i] = 0 \text{ at } u_i^t = v_i \lambda_i^t = -\frac{\dot{r}_{iit}}{E_i[x_i]}[\frac{\lambda_i^t}{\lambda_i^l} + 1]. \tag{6.7}$$

The same way we find $u_i^l = v_i \lambda_i^l = -\frac{\dot{r}_{iit}}{E_i[x_i]}[\frac{\lambda_i^l}{\lambda_i^t} + 1].$ (6.8)

From these relations also follows the representation of the control function $v_i = v_i(t,l), i = 1,2$, which corresponds to the control's form in the initial equations (5.1).

Let us specialize the above control, acting both within the controls definition's domain $G$ and at its border $\partial G$, for example, a square. At these controls in $G$ might exist a geometrical set of points, where the partial derivations of eq.(5.1) get a first kind of the discontinues. For a simplicity, let us consider a monotonous smooth curve $\gamma_5$ (fig.2) as such a set. Generally, such a curve does not cross the above border, and we can prolong this curve by two auxiliary curves $\gamma_2, \gamma_4$ up to $\partial G$ is such a way that the obtained $\gamma_2 \cup \gamma_4 \cup \gamma_5$ will be a monotonous curve (leaving the method of continuation being an arbitrary). As a result of these formations, the initial two bound domain splits on two single subdomains $G_1, G_2$ with the borders $\partial G_1, \partial G_2$ accordingly (fig.2). Because the curve $\gamma_5$ is a priory unknown, the above subdomains are variable.

The following relations formalize the considered domain's and subdomains' descriptions:
$$\bar{G} = G \cup \partial G \cup \gamma_5; G = G_1 \cup G_2 \cup \gamma_2 \cup \gamma_4; \partial G = \gamma_1 \cup \gamma_3 \cup \gamma_6 \cup \gamma_7;$$
$$\bar{G}_1 = G_1 \cup \partial G_1; \partial G_1 = \gamma_1 \cup \gamma_{31} \cup \gamma_2 \cup \gamma_5 \cup \gamma_4 \cup \gamma_{61} \ \bar{G}_2 = G_2 \cup \partial G_2; \tag{6.9}$$

$$\gamma: \begin{cases} t = t_o = s = const \\ l_o \leq l \leq l_k \end{cases}; \ \gamma_3 = \gamma_{31} \cup \gamma_{32}, \ \gamma_{31}: \begin{cases} l = l_o = const \\ t_o = s \leq t \leq t_3^{12} \end{cases};$$

where at $t_3^{12}: \begin{cases} F_2(l,t) = 0 \\ l = l_o \end{cases}$, equation $F_m(l,t) = 0, m = 2,4$ describe any curve in the domain considered below for

$$\gamma_{32}: \begin{cases} l = l_o = const \\ t_3^{12} \leq t \leq t_k \end{cases}; \ \gamma_6 = \gamma_{61} \cup \gamma_{62}, \ \gamma_{61}: \begin{cases} l = l_k = const \\ t_o = s \leq t \leq t_6^{12} \end{cases};$$

where $t_6^{12}: \begin{cases} F_4(l,t) = 0 \\ l = l_k \end{cases}, \gamma_{62}: \begin{cases} l = l_k = const \\ t_6^{12} \leq t \leq t_k \end{cases}; \ \gamma_7: \begin{cases} t = t_k = const \\ l_o \leq l \leq l_k \end{cases}.$ (6.10a)

The border domain has the form
$$\Gamma = \partial G_1 \cup \partial G_2 = \partial G_2 \cup [(\gamma_2^+ \cup \gamma_5^+ \cup \gamma_4^+) \cup (\gamma_2^- \cup \gamma_5^- \cup \gamma_4^-)] = \partial G \cup \Gamma_{int}^+ \cup \Gamma_{int}^-, \tag{6.10b}$$

where $\Gamma_{int}$ is an internal part of domain $\Gamma$, + and – mean the particular curve's movement along the above domains accordingly.

Let us implement the border condition $\delta \Delta S_2 = 0$ using Green's form [23] and the above relations:
$$\iint_{\bar{G}} (\frac{\partial P_1}{\partial l_1} + \frac{\partial P_2}{\partial l_2}) dl_1 dl_2 = \int_{\Gamma} -P_1 dl_2 + P_2 dl_1 \tag{6.11}$$

$$P_1 = \sum_{i=1}^{2} \frac{\partial L}{\partial \dot{x}_{il_1}} \delta x_i - \sum_{i=1}^{2} \frac{\partial L}{\partial \dot{x}_{il_1}} \sum_{i=1}^{2} \frac{\partial x_j}{\partial l_j} \delta l_j + L \delta l_1; P_2 = \sum_{i=1}^{2} \frac{\partial L}{\partial \dot{x}_{il_2}} \delta x_i - \sum_{i=1}^{2} \frac{\partial L}{\partial \dot{x}_{il_2}} \sum_{i=1}^{2} \frac{\partial x_j}{\partial l_j} \delta l_j + L \delta l_2;$$

Applying relations (6.11) to functional (6.1, 6.1a), we come to
$$\delta \Delta S_2 = \iint_{\bar{G}} [\sum_{k=1}^{2} \frac{\partial L}{\partial l_k} \sum_{i=1}^{2} (\frac{\partial L}{\partial \dot{x}_{il_k}} \delta x_i - \sum_{i=1}^{2} \frac{\partial L}{\partial \dot{x}_{il_k}} \sum_{j=1}^{2} \frac{\partial x_j}{\partial l_j} \delta l_j + L \delta l_k)] dl_1 dl_2 = \int_{\Gamma} -P_1 dl_2 + P_2 dl_1$$

Because of the $\delta x_i, \delta l_i$ arbitrariness we get

$$\int_{\partial G} -P_1 dl_2 + P_2 dl_1 = 0 \text{ , (6.12a)} \qquad \int_{\Gamma_{int}^+ \cup \Gamma_{int}^-} -P_1 dl_2 + P_2 dl_1 = 0 \text{ .} \tag{6.12b}$$

The first of them (6.12a) leads to the natural border conditions at the external border of $G$.
For example, in the following forms:

$$\frac{\partial L}{\partial \dot{x}_{it}}\Big|_{t=s} = 0 \Rightarrow h_{ii}(s,l) = 0, x_i \neq 0; \frac{\partial L}{\partial \dot{x}_{it}}\Big|_{t=t_k} = 0, \Rightarrow h_{ii}(t_k,l) = 0, x_i \neq 0;$$

$$\frac{\partial L}{\partial \dot{x}_{il}}\Big|_{l=l_o} = 0, \Rightarrow h_{ii}(l_o,t) = 0, x_i \neq 0; \frac{\partial L}{\partial \dot{x}_{it}}\Big|_{t=s} = 0, \Rightarrow h_{ii}(l_k,t) = 0, x_i \neq 0. \tag{6.13}$$

The second relation (6.12b) leads to an analogy of the Erdman-Weierstrass' conditions [24] at the curve $\gamma_5$. Indeed. Because $\gamma_2, \gamma_4$ are the arbitrary (being the virtual) curves, at crossing them, the partial derivations are continuous, and the integral, taken along the opposite directions, is equal to zero.
From that for (6.12b) is fulfilled

$$\int_{\Gamma_{int}^+ \cup \Gamma_{int}^-} -P_1 dl_2 + P_2 dl_1 = \int_{\gamma_5^+ \cup \gamma_5^-} -P_1 dl_2 + P_2 dl_1. \tag{6.14}$$

Suppose the curve $\gamma_5$ can be defined by equation $l = l^*(t)$.
Then integral (6.14), written in a single (arbitrary) direction, acquires the forms:

$$\int_{\gamma_5} -P_1 dl_2 + P_2 dl_1 = \int_{\tau_1}^{\tau_2} (-P_{11} \dot{l}^* + P_2) dt = 0; \tag{6.15}$$

$$\int_{\tau_1}^{\tau_2} \{\sum_{i=1}^2 (\frac{\partial L}{\partial \dot{x}_{il_2}} \frac{\partial x_i}{\partial l_2} - \dot{l}^* \frac{\partial L}{\partial \dot{x}_{il_1}}) \delta x_i + [L - \sum_{i=1}^2 \frac{\partial L}{\partial \dot{x}_{il_2}} \frac{\partial x_i}{\partial l_2} + \dot{l}^* \sum_{i=1}^2 \frac{\partial L}{\partial \dot{x}_{il_1}} \sum_{i=1}^2 \frac{\partial x_i}{\partial l_2}] \delta l_2 + [-L + \sum_{i=1}^2 \frac{\partial L}{\partial \dot{x}_{il_1}} \frac{\partial x_i}{\partial l_1} \dot{l}^* - \sum_{i=1}^2 \frac{\partial L}{\partial \dot{x}_{il_2}} \sum_{i=1}^2 \frac{\partial x_i}{\partial l_1}] \delta l_1 \} dt = 0.$$

Writing the integral in the opposite directions at the arbitrareness of $\delta x_i, \delta l_i$, we get the system of four equations along $\gamma_5$:

$$h_{ii}^- x_i - \dot{l}^* h_{ii}^- x_i = h_{ii}^+ x_i - \dot{l}^* h_{ii}^+ x_i, h_{ii}^- = h_{ii}^+, i = 1,2, \dot{l}^* = 1, (i=1,2), \tag{6.15a}$$

$$L^- - \sum_{i=1}^2 h_{ii}^- x_i \frac{\partial x_i^-}{\partial l} + \dot{l}^* (h_{11}^- x \frac{\partial x^-}{\partial l} + h_{22}^- y \frac{\partial y^-}{\partial l}) = L^+ - \sum_{i=1}^2 h_{ii}^+ x_i \frac{\partial x_i^+}{\partial l} + \dot{l}^* (h_{11}^+ x \frac{\partial x^+}{\partial l} + h_{22}^+ y \frac{\partial y^+}{\partial l}) \text{ , (6.15b)}$$

where the indexes + and − indicate the functions's values from the domains $G_1$ and $G_2$ accordingly. Substituting (6.15a) into (6.15b) we come to the system of equalities, which determine a jump of the Lagrangian on $\gamma_5$:

$$L^- - L^+ = x(h_{11}^- - \dot{l}^* h_{11}^-)(\frac{\partial x^-}{\partial l} - \frac{\partial x^+}{\partial l}) + y(h_{22}^- - \dot{l}^* h_{22}^-)(\frac{\partial y^-}{\partial l} - \frac{\partial y^+}{\partial l}),$$

$$(L^- - L^+)\dot{l}^* = x(h_{11}^- - \dot{l}^* h_{11}^-)(\dot{x}_t^+ - \dot{x}_t^-) + y(h_{22}^- - \dot{l}^* h_{22}^-)(\dot{y}_t^+ - \dot{y}_t^-). \tag{6.16}$$

The obtained relations are equivalent along the curves

$$\frac{\dot{x}_t^+}{\dot{x}_l^+} = \frac{\dot{x}_t^-}{\dot{x}_l^-} = \frac{\dot{y}_t^+}{\dot{y}_l^+} = \frac{\dot{y}_t^-}{\dot{y}_l^-} = \dot{l}^*; \text{ (6.16a)} \qquad \frac{D^+(x,y)}{D^+(t,l)} = \frac{D^-(x,y)}{D^-(t,l)} = 0, \text{ (6.16b)}$$

which stitch the solutions of (5.1) at the singular surface $\Delta S$ (5.2).
According to (6.16a) the controls at the singular curve become bound:

$$\frac{u_1^t}{u_2^t} = \frac{\dot{r}_{11l}}{\dot{r}_{22l}} \frac{E[x_2]}{E[x_1]} \frac{[\frac{\lambda_1^t}{\lambda_1^l}+1]}{[\frac{\lambda_2^t}{\lambda_2^l}+1]}, \frac{v_1}{v_2} = \frac{\lambda_2^t}{\lambda_1^t} \frac{\lambda_2^l}{\lambda_1^l} [\frac{\lambda_1^t + \lambda_1^l}{\lambda_2^t + \lambda_2^l}] \frac{E[x_2]}{E[x_1]} l^*. \tag{6.17}$$

Let's assume that along the singular curve the conditional probability density is defined by a $\delta$-distribution. Then, according to the features of $\delta$-function, we get the equivalent relations: $\frac{\dot{x}_{it}^{\pm}}{\dot{x}_{il}^{\pm}} = -l^*$ and $\frac{\dot{r}_{11t}^{\pm}}{\dot{r}_{22l}^{\pm}} = l^*$, and the relation for the Lagrangian's jump we write in the form

$$\Delta L = xh_{11}^-(1+l^*)(\dot{x}_l^- - \dot{x}_l^+) + yh_{22}^-(1+l^*)(\dot{y}_l^- - \dot{y}_l^+) = xh_{11}^-(1-\frac{\dot{x}_t^-}{\dot{x}_l^-})(\dot{x}_l^- - \dot{x}_l^+) + yh_{22}^-(1+\frac{\dot{y}_t^-}{\dot{y}_l^-})(\dot{y}_l^- - \dot{y}_l^+) \tag{6.17a}$$

at $h_{ii}^- = h_{ii}^+, i=1,2$.

Because on $\gamma_s$ holds true $x_i = M_i[x_i], \dot{x}_{it}^- = M_i[\dot{x}_{it}^-], \dot{x}_{il} = M[\dot{x}_{il}]$, the following equality is correct:

$$\Delta L = (1+\frac{\dot{r}_{11t}^-}{\dot{r}_{11l}^-})r_{11}^{-1}(\dot{r}_{11l}^- - \dot{r}_{11l}^+) + (1+\frac{\dot{r}_{22t}^-}{\dot{r}_{22l}^-})r_{22}^{-1}(\dot{r}_{22l}^- - \dot{r}_{22l}^+). \tag{6.18}$$

According to (6.16a) and (6.18) we get

$$\frac{\dot{r}_{11t}^{\pm}}{\dot{r}_{11l}^{\pm}} = \frac{\dot{r}_{22t}^{\pm}}{\dot{r}_{22l}^{\pm}} \tag{6.18a}, \Delta L = [1+(\frac{\dot{r}_{11t}}{\dot{r}_{11l}})^{\pm}]r_{11}^{-1}(\dot{r}_{11l}^- - \dot{r}_{11l}^+) + r_{22}^{-1}(\dot{r}_{22l}^- - \dot{r}_{22l}^+). \tag{6.18b}$$

Now we can determine the functional's value at the extremals of equation (5.1):
$$\Delta S = \iint_{\tilde{G}} L dl dt = \iint_{\tilde{G}} -(x\dot{X}_t + x\dot{X}_l + y\dot{Y}_t + y\dot{Y}_l) dl dt + \iint_{\partial G \cup \gamma_5} (-xh_{11}x - yh_{22}y) dl + (xh_{11}x - yh_{22}y) dt = \Delta S_{\text{int}} + \Delta S_{\Gamma}, \tag{6.19}$$

where $\dot{X}_t, \dot{X}_l, \dot{Y}_t, \dot{Y}_l$ are the corresponding covariant functions, $\Delta S_{\text{int}}$ is an internal functional's increment, $\Delta S_{\Gamma}$ is a border's increment, and the Lagrangian is represented by the sum:
$$L = L_1 + L_2 = \dot{x}_t h_{11} x + \dot{x}_l h_{11} x + \dot{y}_t h_{22} y + \dot{y}_l h_{22} y = \dot{x}_t X + \dot{x}_l X + \dot{y}_t Y + \dot{y}_l Y =$$
$$\frac{\partial}{\partial t}(xX + yY) + \frac{\partial}{\partial t}(xX + yY) - x\dot{X}_t - y\dot{Y}_t - x\dot{X}_l - y\dot{Y}_l.$$

According to (6.6) we have

$$\Delta S_{\text{int}} = -\iint_{\tilde{G}} [\sum_{i=1}^{2}(x_i h_{ii} \dot{x}_t + x_i h_{ii} \dot{x}_{il})] dl dt \text{ and } \vec{E}[\Delta S_{\text{int}}] = -\iint_{\tilde{G}} \{\sum_{i=1}^{2} \dot{r}_{iit} + \dot{r}_{iil}\}(\dot{r}_{ii})^{-1}\} dl dt \equiv 0, \tag{6.19a}$$

where $\vec{E} = \{E_1, E_2\}$ is a symbol of mathematical expectation acting additively on the Lagrangian;

$$\Delta \hat{S}_{\Gamma} = \vec{M}[\Delta S_{\Gamma}] = 2\iint_{\partial G \cup \gamma_5} -dl + dt = 2\iint_{\gamma_5} -dl + dt = 2\int_{\tau_1}^{\tau_2}(1-l^*)dt = 2[(\tau_2 - \tau_1) + l^*(\tau_1) - l^*(\tau_2)]. \tag{6.19b}$$

At $l^* = 1$ we get $\Delta S_{\Gamma} = 0$, which brings to a total entropy's increment in the optimal process equals to zero, where $l^* = 1$ corresponds, in particular, the fulfillment of $\dot{x}_t = \dot{x}_l, \dot{y}_t = \dot{y}_l$, i.e. appearance (according to (5.7b,c) a singular curve by the equalization of the above phase speeds. At $l^* > 1$ the entropy's increment is positive, at $l^* < 1$ the increment is negative.

Let us build at an $\varepsilon$-locality of the singular curve a domain $\tilde{G} = \tilde{G}_1 \cup \tilde{G}_2$; $\tilde{G}: \begin{cases} 0 < |l^* - \tilde{l}^*| < \varepsilon \\ t_6^{12} - o(\Delta t) \le t \le t_3^{12} + o(\Delta t) \end{cases}$.

Then the relations (6.7), (6.8) in $\tilde{G}_1$ and $\tilde{G}_2$ holds true, specifically in the forms:

$$\tilde{G}_1 : u_{i1}^t = -\frac{\dot{r}_{ii1l}}{E_i[x_i]}[\frac{\lambda_i^t}{\lambda_i^l}+1], u_{i1}^l = -\frac{\dot{r}_{ii1t}}{E_i[x_i]}[\frac{\lambda_i^l}{\lambda_i^t}+1]; \tag{6.20a}$$

$$\tilde{G}_2 : u_{i2}^t = -\frac{\dot{r}_{ii2l}}{E_i[x_i]}[\frac{\lambda_i^t}{\lambda_i^l}+1], u_{i2}^l = -\frac{\dot{r}_{ii2t}}{E_i[x_i]}[\frac{\lambda_i^l}{\lambda_i^t}+1], \tag{6.20b}$$

where the lover indexes 1,2 at $\dot{r}_{iit}, \dot{r}_{iil}, u_i^t, u_i^l$ indicate that these functions belong to the $\tilde{G}_1$ and $\tilde{G}_2$ accordingly. Using these relations we find the control's jumps:

$$\Delta u_i^t = u_{i1}^t - u_{i2}^t = \lambda_i^t(v_{i1}-v_{i2}) = \frac{(\dot{r}_{ii2l}-\dot{r}_{ii1l})}{E_i[x_i]}[\frac{\lambda_i^t}{\lambda_i^l}+1], \tag{6.21a}$$

$$\Delta u_i^l = u_{i1}^l - u_{i2}^l = \lambda_i^l(v_{i1}-v_{i2}) = \frac{(\dot{r}_{ii2t}-\dot{r}_{ii1t})}{E_i[x_i]}[\frac{\lambda_i^l}{\lambda_i^t}+1]. \tag{6.21b}$$

Therefore, in a general case, there exist the jumps for both the controls and Lagrangian (according to (6.18b) at crossing the singular curve. These jumps can be found if the derivatives of the corresponding correlation functions are known. The conditions $\dot{r}_{iit}^- = \dot{r}_{iil}^-, \dot{r}_{iit}^+ = \dot{r}_{iil}^+$ in particular, for the concentrated systems (at $\dot{r}_{iil}=0$) acquire the forms

$$\dot{r}_{iit}^- = M[\dot{x}(\tau)x^T(\tau+o)], \dot{r}_{iit}^+ = M[\dot{x}(\tau_1)x^T(\tau_1+o)], \tau_1 = \tau+o; \tag{6.22}$$

$$\dot{r}_{iit}^- = \dot{r}_{iit}^+, M_i[x_i(\tau)\dot{x}_i(\tau+o)] + M_i[\dot{x}_i(\tau)x_i(\tau+o)] = 0, x(\tau) = x^-, x(\tau+o) = x^+; x^- = x^+. \tag{6.22a}$$

From that we have $x(\tau) = x(\tau+o)$ and $\dot{x}(\tau) = -\dot{x}(\tau+o)$. Thus, at crossing the singular curve, or a singular point, $\dot{x}(\tau)$ changes sign. If $\dot{x}(\tau) = \lambda(\tau)(x(\tau)+v(\tau))$, $\dot{x}(\tau+o) = \lambda(\tau)(x(\tau)+v(\tau))$ then the control, at crossing the singularity, is found from relation $\lambda(\tau)(x(\tau)+v(\tau)) = -\lambda(\tau)(x(\tau)+v(\tau))$, or $v(\tau) = -2x(\tau)$ and $v(\tau_1) = -2x(\tau_1)$, which determines the needle control (sec.2): $v(\tau)-v(\tau_1) = -\delta v(\tau)$.

The control's strategy that solves the natural border problem consists of:
-the movement along an extremal (6.5) by applying controls (6.7,6.8), being the functions of the initial distribution (5.1a), up to the moment of time when the conditions (6.16a) are fulfilled and the controls become bound by (6.17);
-the movement along a singular curve (at the control's jump) until the condition (6.16a) is violated;
-the movement's continuation along the above extremals with the controls (6.7,6.8).
The following proposition summarizes the results.
*Proposition. The natural border problem's solutions for the path functional with the model (5.1,5.1a) are both the extremal (6.5) and the singular curve of this equation, for which (6.16a) holds true and the controls are bound according to (6.20a,b).*
*Along the singular curve (and/or the singular points) the initial model's dimension is shortening and the state's cooperation takes place.*
All these results follow from the solution of variation problem for the information path functional (sec.2). According to the initial VP, the IPF's extremals hold the principle of stationary action. This allows us to find the invariant conditions, as the model field's functions, being the analogies of the information form of *conservations laws*. Following to the Noether theorem [24] and the results [8] we come to

$$\vec{Q} = [\sum_{i=1}^{n} \frac{\partial L}{\partial(\partial x_i/\partial l_k)}(\sum_{m=1}^{4} -\frac{\partial x_i}{\partial l_m}y_m) + Ly_k]_{k=1}^4 = 0, y_k = \frac{\partial l_k}{\partial t}, l_4 = t. \tag{6.23}$$

Let's have a four dimensional volume $\Omega$ limited by a surface $\Sigma^4$:

$$\Sigma^4 = \Sigma \cup \Sigma_1 \cup \Sigma_2 ; \ \Sigma = \Sigma^3 \cap (l_4 > a) \cap (l_4 < b);$$
$$\Sigma_1 = (F(l_1,l_2,l_3) \leq 0) \cap (l_4 = a); \Sigma_2 = (F(l_1,l_2,l_3) \leq 0) \cap (l_4 = b), \quad (6.24)$$

where $a,b$ are the auxiliary fixed moments of time; $\Sigma^3$ is a no self-crossing surface defined by eqs $F(l_1,l_2,l_3)=0$; $\Sigma^4$ is a four dimensional cylindrical surface limited by two parallel planes $l_4 = a, l_4 = b$, where the cylinder's vertical is in parallel to the time axis and the basis is a geometrical space of points $\Sigma^3$.

After integrating (6.23) by $\Omega$, applying the Ostrogradsky-Gauss theorem [23], we get

$$\int_\Omega div \vec{Q} dv^4 = \int_{\Sigma^4} (\vec{Q}, \vec{n}^+) d\sigma^4 = 0, \quad (6.25)$$

where $\vec{n}^+$ is a positive oriented external normal to the surface $\Sigma^4$; $d\sigma^4$ is an infinite small element of $\Sigma^4$.
Integral (6.25) is represented by the sum of the following integrals, taken by the two cylinder's button parts $\Sigma_1$, $\Sigma_2$ and its sidelong part $\Sigma$ of $\Sigma^4$:

$$\int_{\Sigma^4} (\vec{Q},\vec{n}^+) d\sigma^4 = \int_{\Sigma_1} (\vec{Q},\vec{n}_1^-) d\sigma_1 + \int_{\Sigma_2} (\vec{Q},\vec{n}_1^+) d\sigma_2 + \int_\Sigma (\vec{Q},\vec{n}_2^+) d\sigma = \int_{G^3} [(\vec{Q},\vec{n}_1^+)|_{l_4=b} - (\vec{Q},\vec{n}_1^-)|_{l_4=a}] dv^3 + \int_\Sigma (\vec{Q},\vec{n}_2^+) d\sigma$$
$$= \int_{G^3} [(\vec{Q},\vec{n}_1^+)|_{l_4=b} - (\vec{Q},\vec{n}_1^+)|_{l_4=a}] dv^3 + \int_\Sigma (\vec{Q},\vec{n}_2^+) d\sigma, \quad (6.26)$$

where $\vec{n}_1^+ = (0,0,0,1)$ is a positive oriented external normal to the bottom part $\Sigma_2$ of the surface $\Sigma^4$; $l_4 = t$; $\vec{n}_1^- = (0,0,0,-1) = -\vec{n}_1^+$ is a negative (internal) normal to the bottom part $\Sigma_1$ of $\Sigma^4$; $\vec{n}_2^+ = (\vec{n}_{21}^+, \vec{n}_{22}^+, \vec{n}_{23}^+, 0)$ is a positive external normal to $\Sigma$; $G^3 = (F(l_1,l_2,l_3) \leq 0) \cap (l_4 = 0)$ is a part of space being a projection of $(\Sigma_1, \Sigma_2)$ on a hyper plane $l_4 = 0$; $dv^3$ is an infinite small element of volume $G^3$; $d\sigma$, $d\sigma_1$, $d\sigma_2$ are the infinite small elements of the surfaces $\Sigma, \Sigma_1, \Sigma_2$ accordingly.

Let us implement (6.26) at the usual physical assumptions, supposing that both the function $F(l_1,l_2,l_3) = l_1^2 + l_1^2 + l_1^2 - R_o^2$ and the field are decreasing fast on infinity.
This means that at $R_o \to \infty$ and $d\sigma \sim R_o^2$, the integral by $\Sigma$ in (6.26) can be excluded.
Then (6.26) according to (6.25) acquires the form

$$\int (\vec{Q},\vec{n}_1^+)|_{l_4=b} dv^3 = \int (\vec{Q},\vec{n}_1^-)|_{l_4=a} dv^3 \quad (6.27)$$

where the integral is taken by an infinite domain. Because of the auxiliary $a$ and $b$, the above equality means the preservation in time the values

$$\int (\vec{Q},\vec{n}^+) dv^3 = \int [\sum_{i=1}^n \frac{\partial L}{\partial(\partial x_i / \partial l_k)}(\sum_{m=1}^4 -\frac{\partial x_i}{\partial l_m} y_m) + L] dv^3, \ y_m = \frac{\partial l_m}{\partial t}, l_4 = t. \quad (6.28)$$

Applying the Lagrange-Hamilton equations we get invariant

$$\int [\sum_{i=1}^n (\frac{\partial L}{\partial(\partial x_i / \partial t)} \frac{\partial x_i}{\partial t}) + L] dv^3 = \int \sum_{i=1}^n (X_i \frac{\partial x_i}{\partial t} + L) dv^3 = \int (H + 2L) dv^3 = inv, \ dv^3 = dl_1 dl_2 dl_3, \quad (6.29)$$

which at $-H = L - \dot{x}^T X, L = 1/2 \dot{x}^T X$ (sec.2) leads to the invariant

$$\int H dv^3 = inv, H = 1/2 \sum_{i=1}^n (\frac{\partial x_i}{\partial \vec{l}}, \frac{\partial \vec{l}}{\partial t}) X_i, \quad (6.30)$$

preserving the volume's Hamiltonian of the information path functional.

# 7. The connection between the entropy's (information) path functional (IPF) and the Kolmogorov's (K) entropy of a dynamic system, between the Kolmogorov's and the macrodynamic complexities, and the relations to physics

The K-entropy is an entropy per unit of time, or the entropy production rate, measured by a sum of the Lyapunov's characteristic exponent (LCE) [5, 25-27]. LCE describes a separation between the process' trajectories, created by the process dynamic peculiarities.

In the IPF model, the separation is generated by the inner controls actions, which carries out the transitions between the process' dimensions, physically associated with the phase transformations, singularities, chaotic movement and related physical phenomena [28-31].

Let us find the LCE for the IPF model. At the DP, each of these controls swiches the process' extremal segment (with an eigenvalue $\lambda_i$) *from* a local movement $x_{it} = x_{io} \exp(-\lambda_i t)$, corresponding a local process' stability, *to* the local movement $x_{i\tau} = x_{i\tau o} \exp(\lambda_{i\tau} t), t \in (\tau - o, \tau)$, (7.1)
corresponding a local process' instability, which brings a separation between these two process' movements. Here $x_{io}$ is an initial condition at a beginning of the *i*-segment; with the macroprocess' eigenvalue $\lambda_i$, $x_{i\tau o}$ is a starting state at the moment $t = \tau - o$ (near the segment end), $\lambda_{i\tau}$ is the eigenvalue at $\tau - o$ approaching $\tau$ (which depends on $gradX(\tau - o)$, sec.2) that potentially initiates these dynamics, approximating the between segment's stochastics at $t \to \tau$).

The LCE is measured by a mean rate of exponential divergence (or convergence) of two neighboring trajectories: one of them describes an initial undisturbed movement $x_{it}$, another one is the disturbed movement $x_{i\tau}$ (for this model at DP). A local LCE :

$$\sigma_i = \lim_{t \to \tau} \frac{1}{t} \ln(\frac{x_{i\tau}}{x_{i\tau o}}) = \lambda_{i\tau} \qquad (7.1a)$$

expresses the exponential divergence $x_{i\tau}$ from the movement $x_{it}$ along the extremal segment ($x_{i\tau}$ starts at the moment $t = \tau - o$ by the end of the movement $x_{it}|_{t=\tau-o} \to x_{i\tau o}$, which precedes the beginning of the disturbed movement $x_{i\tau}$). At $\lambda_{i\tau} > 0$, the process is instable and chaotic: the nearby points, no matter how they close, will diverge to any arbitrary separation. These points are instable. At $\lambda_{i\tau} < 0$, the process exhibits asymptotic stability in a dissipative or a non-conservative system. The LCE zero at $\lambda_{i\tau} = 0$ indicates that the system is in a steady state. A physical system with this exponent is a conservative. Such a system exhibits Lyapunov stability. Although this system is deterministic, there is no process' order in this case [30, 31].

Exponent (7.1) approximates the dynamic divergence of the extremal segments at a window between the segments; and the LCE (7.1a) characterizes the *information dynamic* peculiarities arising at the DP localities. In particular, under the optimal control, applied to $\lambda_{i\tau}$ at the nearest moment $\delta\tau$ following $\tau$, the eigenvalue changes according to equations

$$\lambda_{i\tau}(v_{i\tau}) = -\lambda_{i\tau} \exp(\lambda_{i\tau}\delta\tau)[2 - \exp(\lambda_{i\tau}\delta\tau)]^{-1}, \text{ which at } \delta\tau \to 0 \text{ reaches a limit: } \lim_{\delta\tau \to 0} \lambda_{i\tau} = -\lambda_{i\tau}.$$

Such a discrete (jump-wise) LCE renovation, is a phenomenon of a controllable process, specifically at the process' coupling, and could serve as a LCE indicator of this phenomenon.

The K entropy is the nonlinear dynamics counterpart of physical the Boltzmann-Gibbs entropy [32], which is directly connected to the Shannon's information entropy [21].

The IPF model's DPs are the crucial points of changes in a *dynamical evolution* with the fixed entropy path functional's production rates (PFR), given by the sum of positive LCE.

According to relation (2.36), the PFR, being equal to the sum of the operator's positive eigenvalues:

$$E[-\frac{\partial S^i}{\partial t}(\tau)] = E[H(\tau)] = Tr[A(\tau)] = \sum_{i=1}^{n} \lambda_i(\tau_i) > 0, \quad (7.1b)$$

coincides with the K entropy at these crucial points. In the analogy between statistical mechanics and chaotic dynamics. This additivity of the discrete linear rate (at DPs) for both the K entropy and PFR corresponds to a thermodynamic extensivity of the Boltzmann-Gibbs entropy [33], which is important in a connection between statistical mechanics and chaotic dynamics. The extensivity of entropy is an essential requirement with which thermodynamics can be created [33-35]. This may be the case even if a system energy is nonextensive [34].

A sufficient important is the linear growth of the K entropy and the thermodynamic extensivity of the Boltzmann-Gibbs entropy only in the long-time limit and the thermodynamic limit, respectively. As it's known [33], a physical quantity to be a temporally extensive should satisfy its linear grow in time. Thus, for example, the K entropy possesses the temporal extensivity for chaotic dynamical systems.

The IPF model holds the open system's qualities such as a nonlinearity and irreversibility (at the DP), and the stationarity and reversibility within each extremal segment, corresponding a system's conservativity. These phenomena allow applying the IPF model for a wide class of real systems, which show the above alternative behaviors at different stages of dynamic evolution [36,37].

Most publications on this subject are based on the models of the linear phenomenological irreversible thermodynamics, using an energetic approach, fluctuation from a stationary state, or a quasi equilibrium process [38-41]. Actual irreversible macroprocess might arise from a random movement at microlevel with a random entropy, using the information approach, while the relations of preservation energy could not be fulfilled.

The main problem consists of math difficulties of applying a *macro* evolution approach to a random process and random entropy. Some publications use an informational approach to self-organization, applying a control's *parameter* for an evaluation of irreversibility in a state's transition [42]. The equations for a *controllable* irreversible information macroprocess are still unknown. The VP, applied to information path functional, defined on the solution of a controllable stochastic equation, brings the *irreversible kinetic macroequation* and its connection with *diffusion*. Applying the Shannon's entropy measure for a multi-dimensional random *process* with the statistical dependent events leads to the unsolved problem of the long terms $n$-dimensional correlations, while these events are a *naturally connected* by the entropy path functional.

The lack of additivity—even for statistically independent events—leads to the problem related to the lack of thermodynamic extensivity [35]. The entropy measure of degree-α and the α-norm entropy measure [43-44] satisfy a "pseudo-additive" relation, associated with a *nonextensive* thermodynamics, rather than the additive relation, provided by the Shannon and Renyi [45] entropies.

The evolutionary path functional's entropy is defined by a simple sum of the local entropies at each DP, according to (7.1b) that is applied to an extensive dynamic system. But the extensivity is locally violated at the random window between the extremal segments. The evolutionary PFR forms a ranged sum, satisfying the VP. The maximal and minimal PFR values characterize the maximal and minimal speeds of evolutionary process according to [6]. A current PFR is defined by a sequential enclosure each of a previous model's eigenvalue to the following one, connected by the IN structure. This allows getting the cooperative complexity for all process [7], as well as the PFR measure at each stage of evolution. The IN final node's eigenvalue characterizes both the system's terminal evolutionary speed and the system's *cooperative* complexity [7,8].

Algorithmic Kolmogorov's (K) complexity [5] is measured by the relative differential entropy of one object (*k*) with respect to other object (*j*), which is represented by a shortest program in bits. The *common* entropy measure connects both the K-complexity and the information macrocomplexity $\Delta MC_{kj}^{\delta}$ [7]. So, the $\Delta MC_{kj}^{\delta}$ complexity measures the quantity of information (transmitted by the relative information flow), required to join the object *j* with the object *k*, which can be expressed by the algorithm of a minimal program, encoded in the IN communication code (sec.4). This program also measures a "difficulty" of obtaining information by *j* from *k* in the transition *dynamics*. Assigning a common *digital* $\Delta MC_{kj}^{\delta}$ measure to all communicated objects allows also determining the unknown constant $C_h$ in the K complexity [5]. The $\Delta MC_{kj}^{\delta}$ maximum represents the information measure between *order and disorder* in stochastic dynamics and it can detect determinism amongst the randomness and singularities. Because the IPF has a limited time length, and the IPF strings are finite, being an upper bound, the considered cooperative complexity is *computable* in an opposite to the Kolmogorov's *incomputability. The MC-complexity is able to implement the introduced notion and measure of information independent on the probability measure by applying the IN's information code.*

The above results lead to a mutual connection of the model's *Uncertainty, Regularity, and Stability.*

Uncovering of the regular causes of a random process lays in a foundation of revealing the process regularities. Such an opportunity provides the IPF whose information invariant encodes a chain of regular events, covered by the random process's IPF. Therefore, the IPF measures the process's uncertainty by the entropy functional and allows minimizing them by the applied optimal controls. The IPF's Hamiltonian that determines both an instant entropy production and the process macromodel's operator, also defines the LCE as Lyapunov's function of the process' stability, which connects the stability to the process uncertainty.

The process' optimization by the controls' actions changes the LCE sign at the DP bringing the cooperative process' stability concurrently with the minimization its uncertainty.

*Because most natural processes are random, understanding of their regularities involves the minimization of the random uncertainties by VP, which leads to imposing the dynamic constraints (sec.1) and getting the process information dynamic model with all above peculiarities. That is why the constraint imposition is considered as a general method of revealing dynamic regularities of the random process and its dynamic equations.*

In the considered Maxwell demon's feedback [46], an observer first *transforms* a *random uncertainty* (events) in information (certainty, for example, expressed by specific probability-that is a *non-random measure of the random events*), while information itself is a *non-material substance*. Second, to *acquire* this information and transform it through a feedback, the observer *binds* this information with a source of energy, which is only a *currier* of information. Other carriers are different physical materials (for example, photosensitive elements, etc.). Therefore, the described experiment shows *not* that information contains energy, but the experiment just *evaluates* the energy, which the measurement devices (including sensory, brain) *spend for binding* the incoming information for it *transmission*.

Actually, the *transformation* of random uncertainty in information by a *physical* observer is accompanied with *binding* it with the observer's energy and/or its material substance, which serves as a carrier for the information transmission. The considered VP is a *formal mathematical* mechanism transforming the entropy functional uncertainty to the path functional (IPF) information. According to VP, this transformation requires spending a certain *invariant* quantity of information, which each IPF extremal's segment (a discrete interval) binds (sec.4). The VP defined invariants $\mathbf{a}_o, \mathbf{a}$ take the values 0.70-0.23, depending on the interval length, where the minimum belong to an interval of a delta-function, or generally: $0 \leq a \leq \ln 2$. At a fixed interval, the specific invariants' values evaluate an efficiency of binding information, being concealed within the interval. According to the VP, each invariant measures an extreme quantity of that information which depends on the ratio of imaginary and real eigenvalue for this interval.

Following the formula [46] exponential form $\exp[\langle \Delta F - W \rangle / k_B T] = I$, where $\Delta F$ is a free-energy difference between states, $W$ is a work done on the system, $I$ is information, $T$-temperature, $k_B$ Bolzman constant, and $\langle \bullet \rangle$ is the average of the considered ensemble, we get the same formula in logarithmic form $\langle \Delta F - W \rangle / k_B T = \ln I$, where the energy-bind information takes the value $I \geq 1$, while at $\langle \Delta F - W \rangle = 0, I = 1$.

Therefore, information is binding only at $||\langle \Delta F - W \rangle|> 0, \ln I > 0$.

According to [46], $1 \leq I \leq 2$, which corresponds to $0 \leq \ln I \leq \ln 2$ and it coincides with the VP invariants.
These confirm both the initial concept that information is not an energy, but it rather the energy binds information defined by the VP, and this information is precisely evaluated by the VP invariants.

**Figures**

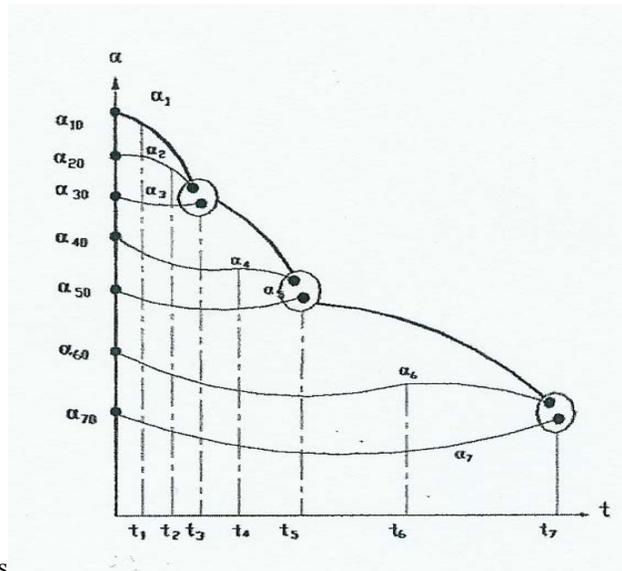

Fig.1. The cooperation of the model's eigenvalues

Fig.1a. A triplet's information structure with applying both the regular and needle controls.

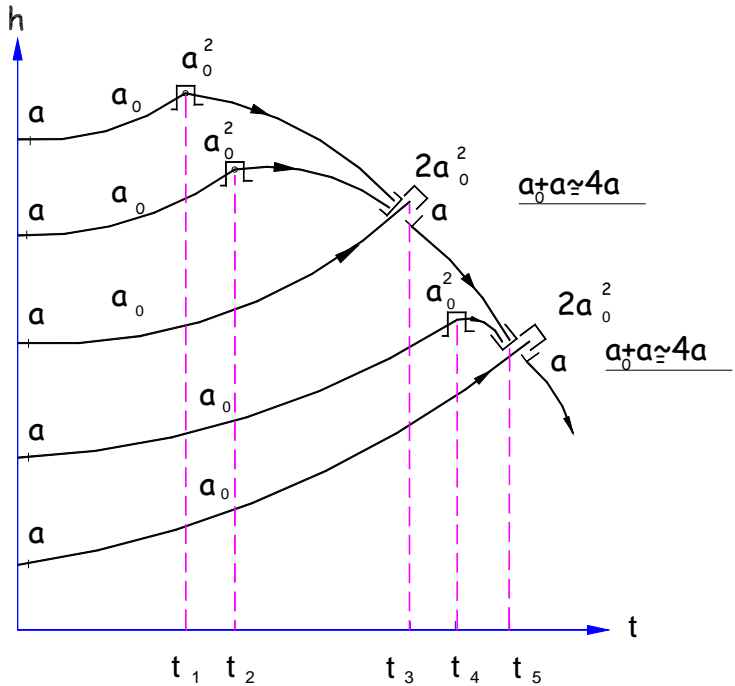

Fig.2. An illustration of the control's domain and the auxiliary curves

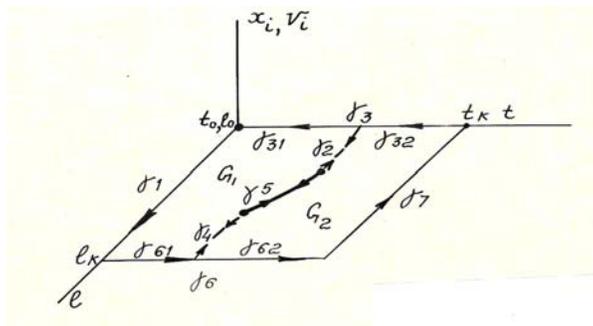